\documentclass[seceq]{ptptex}
\usepackage{graphicx}
\usepackage{bm}
\usepackage{color}
\usepackage{wrapft}

\newcommand{\simle}
{\raisebox{-0.75ex}[-1.5ex]{$\;\stackrel{<}{\sim}\;$}}

\def\rd{{\partial}}
\def\sg{{\sigma}}
\def\eps{{\varepsilon}}
\newcommand{\sxy}{\sigma_{\rm SH}}
\def\bk{{ {\bm k} }}

\def\w{{\omega}}
\def\a{{\alpha}}

\def\g{{\gamma}}
\def\G{{\Gamma}}
\def\ld{{\lambda}}
\def\th{{\theta}}
\newcommand{\eq}{eqnarray}
\newcommand{\f}{\frac}
\def\nn{{\nonumber}}
\newcommand{\aH}{\alpha_{\rm SH}}
\def\ls{{\langle {\bm l} \cdot {\bm s} \rangle_{\mu}}}
\def\lsm{{\langle {\bm l} \cdot {\bm s} \rangle_{-}}}
\def\lsp{{\langle {\bm l} \cdot {\bm s} \rangle_{+}}}
\def\th{{\theta}}
\def\sg{{\sigma}}
\def\V2{{|V_{\bk M \sg}|^2}}

\pubinfo{Vol.~128, No.~5, November 2012}

\title{Extrinsic Spin Hall Effect
Due to Transition-Metal Impurities
}

\author{
Takuro \textsc{Tanaka}
and Hiroshi \textsc{Kontani}
E-mail: kon@s.phys.nagoya-u.ac.jp}

\inst{
Department of Physics, Nagoya University,
Furo-cho, Nagoya 464-8602, Japan. 
}

\recdate{
Recived: August 29, 2012}

\abst{We investigate the extrinsic spin Hall effect in the electron gas model due to
transition-metal impurities based on the single-impurity Anderson model with orbital degrees of freedom.
Both the skew scattering and side jump mechanisms are analyzed in a unified way,
and the significant role of orbital degrees of freedom are clarified. 
The obtained spin Hall conductivities
are in proportion to the 
spin-orbit polarization at the Fermi level 
$\langle {\bm l}\cdot {\bm s} \rangle_{\mu}$ as is the case with the intrinsic spin Hall effect:
skew scattering term $\sigma_{\rm SH}^{\rm ss} \propto \langle {\bm l}\cdot 
{\bm s} \rangle_{\mu} \delta_1\sigma_{xx}$, and side jump term $\sigma_{\rm SH}^{\rm sj} \propto \langle {\bm l}\cdot {\bm s} \rangle_{\mu}$, where $\delta_1$ is the phase shift for $p$ ($l=1$) partial wave.
Furthermore,
the present study indicates the existence of a nontrivial close relationship 
between the intrinsic term $\sxy^{\rm{int}}$
and the extrinsic side jump term $\sxy^{\rm{sj}}$. 
}

\title{
Extrinsic Spin Hall Effect
Due to Transition-Metal Impurities 
}

\author{
Takuro Tanaka and Hiroshi Kontani}

\inst{
Department of Physics, Nagoya University,
Furo-cho, Nagoya 464-8602, Japan. 
}

\date{Received: August 29, 2012}

\abst{
We investigate the extrinsic spin Hall effect in the electron gas model due to
transition-metal impurities based on the single-impurity Anderson model with orbital degrees of freedom.
Both the skew scattering and side jump mechanisms are analyzed in a unified way,
and the significant role of orbital degrees of freedom are clarified. 
The obtained spin Hall conductivities
are in proportion to the 
spin-orbit polarization at the Fermi level 
$\langle {\bm l}\cdot {\bm s} \rangle_{\mu}$ as is the case with the intrinsic spin Hall effect:
skew scattering term $\sigma_{\rm SH}^{\rm ss} \propto \langle {\bm l}\cdot 
{\bm s} \rangle_{\mu} \delta_1\sigma_{xx}$, and side jump term $\sigma_{\rm SH}^{\rm sj} \propto \langle {\bm l}\cdot {\bm s} \rangle_{\mu}$, where $\delta_1$ is the phase shift for $p$ ($l=1$) partial wave.
Furthermore,
the present study indicates the existence of a nontrivial close relationship 
between the intrinsic term $\sxy^{\rm{int}}$
and the extrinsic side jump term $\sxy^{\rm{sj}}$. 
}

\begin{document}
\maketitle


\maketitle

\section{\label{sec1} Introduction}

There is growing interest in the spin Hall effect
(SHE), which is the phenomenon that an electric field induces
a spin current in a transverse direction.
After its theoretical prediction by Hirsch
in 1999 \cite{Hirsch}, which was originally proposed by D'yakonov
and Perel' in 1971 \cite{Perel}, investigations were initially 
focused mostly in semiconductors \cite{Kato}.
Recently, however, experiments have made great progress on
electrical spin injection and detection in metals
\cite{Saitoh,Valenzuela,Kimura}.
Therefore, SHE in various metals has received considerable attention owing 
to its fundamental interest as well as its potential application in spintronics.


The intrinsic SHE arises from the Berry phase of the multiband Bloch function \cite{Sinova-SHE, Murakami-SHE}.
In $d$-electron systems, the spin Hall conductivity (SHC) takes a large value
due to the ``orbital Aharonov-Bohm phase factor" induced by the $d$-angular
momentum with the aid of the atomic spin-orbit interaction (SOI) \cite{Kontani-Ru, 
Kontani-Pt,Guo-Pt, Tanaka-4d5d, Kontani-OHE}.
In Ref. \citen{Kontani-OHE},
the present authors pointed out, for the first time, that
the intrinsic SHC is in proportion to the spin-orbit polarization at the Fermi level $\ls$.
In accordance with the sign change of $\ls$,
SHC changes its sign.
The sign of the spin Hall angle (SHA $\alpha_{\rm{SH}}=\f{j^S_{x}}{j^C_{y}}\f{2|e|}{\hbar}$) calculated in this study
is consistent with the recent experimental observation 
based on the spin pumping method \cite{Hoffmann}.
Furthermore, more recently, SHC experimentally measured 
by Morota {\it et al.} \cite{Morota}
is semiquantitatively consistent with the result in Ref. \citen{Tanaka-4d5d}.
This fact strongly indicates that the intrinsic mechanism is dominant
for the experimentally observed SHE in 4$d$ and 5$d$ transition metals. 
In $f$-electron systems, moreover, which possess larger angular momentum,
a giant intrinsic contribution arises from the same mechanism as $d$-electron
systems \cite{Kontani-df,Tanaka-PAM}. 

In addition to the intrinsic mechanism, SHE is also caused by impurity scattering
with the aid of the SOI, which is known as extrinsic SHE \cite{Perel,Hirsch,Takahashi,Smit,Berger,Bruno}. 
A controversial study on the extrinsic SHE in Au/FePt was reported in Ref. \citen{Mitani}.
To understand this experimental observation,
Guo {\it et al.} studied the skew scattering mechanism 
due to Fe impurities in Au \cite{Nagaosa-FeAu}.
Therein, they considered that the giant SHE is originated from 
the orbital-Kondo effect of Fe impurities in Au host.
However, Mihajlovic {\it et al.} reported the small SHA in Au
for a 60 nm-thick Au Hall bar \cite{negative}.  
Consistently with the latter reported, $\ls$ of Fe will be small since the orbital magnetic moment is almost quenched because of the small SOI of Fe atom \cite{Costi,Shick,Brewer}.
Independently of Ref. \citen{Nagaosa-FeAu}, 
the present authors studied the skew scattering term  due to 
the $f$-electrons \cite{Tanaka-RE}.
Therein, we found that the giant SHC is realized only in the case of
a large spin-orbit polarization.

More recently,
Seki {\it{et al.}} \cite{Seki} and Sugai {\it{et al.}} \cite{Sugai} have studied 
the Au layer thickness dependence of the SHA, and concluded that the giant SHA originates from the the scattering on the surface. 
Later, Gu {\it et al.} studied the possibility for a surface-assisted 
skew scattering on Pt impurities \cite{Gu2}.
In another group, in addition, Niimi and Fert et al. have recently reported a large SHA in Cu due to Ir impurity \cite{Niimi, Fert-PRL}.

In the extrinsic mechanism, not only the skew scattering mechanism but the side jump mechanism also exists.
Compared with the skew scattering mechanism,
the side jump mechanism is less well understood.
To elucidate the major role of the orbital degrees of freedom in the extrinsic SHE,
studies on both the skew scattering and side jump terms based on the single-impurity Anderson model (SIAM) are highly required.
Moreover, the side jump term originates from the anomalous velocity, as is the case with the intrinsic term.
This fact could indicate the existence of a close relationship between the side jump term and the intrinsic term \cite{Lyo}.

In this paper, we study the extrinsic SHE based on the SIAM for transition metal atoms. 
The analytical expressions for both the skew scattering term $\sxy^{\rm{ss}}$ 
and side jump term $\sxy^{\rm{sj}}$ are derived. 
As is the case with the intrinsic term $\sxy^{\rm{int}}$, 
both $\sxy^{\rm{ss}}$ and $\sxy^{\rm {sj}}$ are proportional to the spin-orbit
polarization at the Fermi level $\ls$:
$\sxy^{\rm ss} \propto \langle {\bm l} \cdot {\bm s} \rangle_{\mu} \delta_1 \sigma_{xx}$,
and $\sxy^{\rm sj} \propto \langle {\bm l} \cdot {\bm s} \rangle_{\mu}$, 
where $\delta_1$ is the phase shift for $p$ ($l=1$) partial wave.
Therefore, $\langle {\bm l} \cdot {\bm s} \rangle_{\mu} \sim O(1)$ will be
a necessary condition for a giant extrinsic SHE, which is realized in 
the rare-earth atoms.
In contrast, $\langle {\bm l} \cdot {\bm s} \rangle_{\mu}$ is small 
in 4$d$ and 5$d$ transition metals \cite{Kontani-OHE}, which 
makes the extrinsic SHE small.
Both the intrinsic and extrinsic terms can be discussed from 
the same viewpoint that is the spin-orbit polarization at 
the Fermi level $\ls$.

In the present study, it is found that the contribution of the side jump term
due to the impurity such as Pt in Cu host
is comparable to that of the intrinsic term in Pt. 
In contrast, the condition for the skew scattering term to be huge is difficult to be realized,
which not only $\ls$ but also $\delta_1$ should be large.
\section{\label{sec2} Model and Hamiltonian}

\begin{figure}[!htb]
\includegraphics[width=.4\linewidth,height=.25\linewidth]{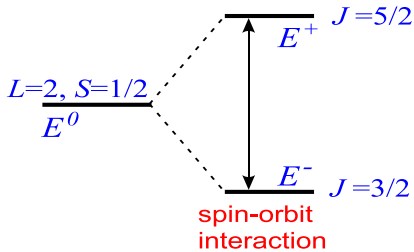}
\caption{\label{fig:split} The atomic SOI $\ld {\bm l}\cdot{\bm s}$ leads to a splitting between $d$-level
with total angular momentum $J=3/2$ and $J=5/2$, whose energies are given 
by $E^- = E^0 -3\ld/2$ and $E^+ = E^0 + \ld$, respectively.
Here, $E^0$ is the $d$-electron energy without SOI.  
} 
\end{figure}

In this study, we use the SIAM for 
transition metal atoms.
In the presence of the SOI, $d$-electron states are specified by the total
angular momentum $J=2\pm 1/2$ and its $z$-component.
By using the eigenvalues for $ \hat J^2, \ \hat L^2$, and $\hat s^2$, we obtain ${\bm l}\cdot {\bm s} = \{ j(j+1) - l(l+1) -s(s+1) \}/2 = -3/2 \ (1)$ for $J=3/2$ ($J=5/2$). 
As shown in Fig.\ref{fig:split}, therefore, $E^{-} = E^0 - \f{3}{2} \ld$ and $E^{+} = E^0 + \f{3}{2}\ld$, where 
$E^0$ is the $d$-level energy without SOI.
In contrast to the rare-earth atoms,
$J=3/2$ and $J=5/2$ states
cannot be treated separately in 4$d$ and 5$d$ metal atoms 
since the $s$-$d$ hybridization potential is larger than the SOI.
Therefore, we introduce the following SIAM
for 4$d$ and 5$d$ transition metal atoms with both $p$- and $d$-orbitals \cite{Fert}:
\begin{eqnarray}
H&=& \sum_{\bk,\sg} \eps_{\bk} c^{\dagger}_{\bk \sg} c_{\bk \sg}
+ \sum_{\bk\sg m} E^p p^{\dagger}_{\sg m} p_{\sg m} + \sum_{\bk,M, \a=\pm } E^{\a} \left( d^{\a}_M \right)^{\dagger} d^{\a}_M \nn \\
&+&\sum_{\bk\sg m} \left\{ V^{p}_{\bk m} c^{\dagger}_{\bk \sg} p_{\sg m} + {\rm h.c.} \right\} 
+ \sum_{\bk,\sg, M, \a=\pm} \left\{ V^{\a}_{\bk M \sg} c^{\dagger}_{\bk \sg} d^{\a}_M + \rm {h.c.} \right\} + \f{U^d}{2} \sum_{M \neq M'} n^d_M n^d_{M'}.  \nn \\ \label{eq:Ham}
\end{eqnarray}
Here, $c^{\dagger}_{\bk \sg}$ is the creation operator of a conduction 
electron with spin $\sg=\pm 1$. 
$\left( d^{\pm}_M \right)^{\dagger}$ is the creation operator of a $d$-electron 
with total angular momentum $J= 2\pm 1/2$ and $z$-component 
$M \ (-J \leq M \leq J )$. 
$p^{\dagger}_{\sg m}$ is the creation operator of a $p$-electron with 
angular momentum $m(-1 \leq m \leq 1 )$.
%
%
$\eps_{\bk}=k^2/2m$ is the energy for the conduction electrons,
 and $E^{\pm}$ ($E^p$) is the localized $d \ (p)$-level energy. 
Here, we omit the atomic SOI for $p$-electrons since it is much smaller than
$|\mu - E^p|$.
$V^{\pm}_{\bk M \sg}$  and $V^{p}_{\bk m}$ are the mixing potentials, which are given by
\begin{\eq}  
V^{\pm}_{\bk M \sg} &=& \sqrt{4\pi} V_d \sum_{m} a^{M\pm}_{m\sg} Y^m_2 (\hat \bk),\\
V^{p}_{\bk m} &=& \sqrt{4\pi} V_p Y^m_1 (\hat \bk),
\end{\eq}
where $a^{M\pm}_{m \sg}$ is the Clebsch-Gordan (C-G) coefficient for $J=2\pm1/2$ and 
$Y^m_l(\hat \bk)$ is the spherical harmonic function.
We will show that the phase factor in $Y_l^m(\hat \bk)$ and 
$a^{M \pm}_{m\sg}$ are indispensable to realize the SHE. 
Here, the C-G coefficient is given by 
\begin{\eq}
a^{M-}_{m\sg} &=& -\sg \left\{ \left( 5/2-M\sg \right)/5  \right\}^{1/2} \delta_{m,M-\sg/2} \ \ \ \text{for} \ \ J=3/2, \nn \\
a^{M+}_{m\sg} &=& \left\{ \left( 5/2+M\sg \right)/5  \right\}^{1/2} \delta_{m,M-\sg/2} \  \qquad \text{for} \ \ J=5/2. \label{eq:J5/2}
\end{\eq}
%
In the present study, we neglect the crystalline electric field for each of the $J=3/2$ and $J=5/2$ states
to avoid too complicated expressions.
We put $\hbar = 1$ hereafter.

%

To discuss the scattering problem, it is useful to
derive an effective Hamiltonian for the conduction electrons
by integrating out the 
$d$ and $p$ electrons in Eq. (\ref{eq:Ham}) \cite{YY,Haldane}.
The obtained Hamiltonian is given by
\begin{\eq}
H_c&=& \sum_{\bk \sg} \eps_{\bk} c^{\dagger}_{\bk \sg}c_{\bk \sg} 
+ \sum_{\bk,\bk',\sg} J^p_{\bk,\bk'} c^{\dagger}_{\bk \sg} c_{\bk' \sg} \nn \\
&+& \sum_{\bk, \bk',\sg,\sg',\a=\pm} J^{\a}_{\bk\sg,\bk'\sg'} c^{\dagger}_{\bk\sg}c_{\bk' \sg'},
\label{eq:Ham-re}
\end{\eq}
where
\begin{\eq}
J^p_{\bk,\bk'} &=& \f{1}{\mu-E^p}\sum_{m} V^p_{\bk m} (V^{p}_{\bk' m})^{\ast} \nn \\
&=& 4\pi J_p \sum_{m} Y^m_1(\hat \bk) \left[ Y^m_1(\hat \bk') \right]^{\ast},  \\ 
J^{\pm}_{\bk\sg,\bk'\sg'} &=& \f{1}{\mu-\tilde E^{\pm}}\sum_{M} V^{\pm}_{\bk M\sg} \left( V^{\pm}_{\bk' M\sg'} \right)^{\ast} \nn \\
&=&4\pi J_{\pm } \sum_{M m m'} a^{M\pm}_{m\sg} a^{M\pm}_{m' \sg'} Y^m_2 (\hat \bk) \left[ Y^{m'}_2(\hat \bk') \right]^{\ast},
\end{\eq}
$ J_p \equiv |V_p|^2/(\mu-E^p)$ and $ J_{\pm} \equiv |V_{d}|^2/(\mu-\tilde E^{\pm})$.
Here, $\tilde E^{\pm} = E^{\pm} + \rm {Re}\it{\Sigma}^{d}$; $\it{\Sigma}^d$ is the $d$-electron 
self-energy due to the Coulomb interaction $U^d$.
\begin{figure}[!htb]
\includegraphics[width=0.7\linewidth,height=.2\linewidth ]{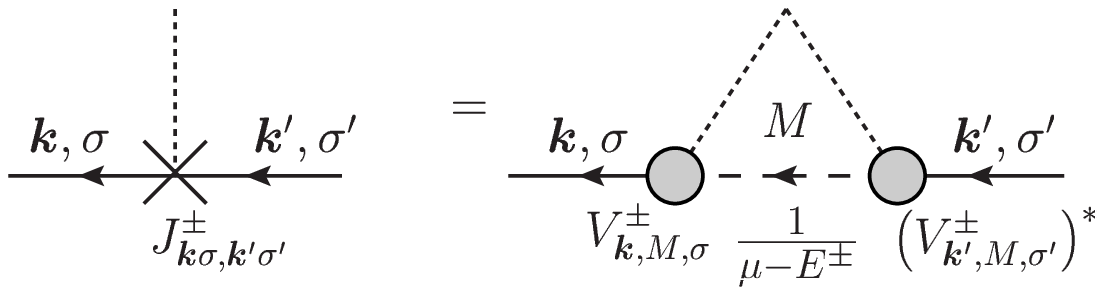}
\caption{\label{fig:sj-tmat} Diagrammatic expression for 
$J^{\pm}_{\bk\sg,\bk'\sg'}$.
} 
\end{figure}
According to the scaling theory \cite{Haldane}, $\tilde E^{\pm} $ approaches the Fermi level as the temperature decreases due to the Kondo effect:
$|J_{\pm}|$ is strongly enhanced near the Kondo temperature $T_{\rm{K}}$, 
and below $T_{\rm K}$,
$J_{\pm} N(0) \gg 1$ due to strong resonant scattering,
where $N(0)=m k_{F}/2\pi^2$ is the density of state of the conduction band per spin. 
We also assume that $J_p N(0) (=-\tan\delta_1/\pi) \ll 1$ 
since $|\mu - E^p | \sim O(1 {\rm eV})$ \cite{Fert}. 


Here, we explain the spin-orbit polarization due to the SOI, which plays 
a significant role in the intrinsic SHE \cite{Kontani-OHE}.
Using the relation $\langle {\bm l}\cdot {\bm s} \rangle_{+(-)} = 1\ (-3/2)$,
the spin-orbit polarization ratio of $d$-electrons at the Fermi level is given by
\begin{\eq}
\langle {\bm l} \cdot {\bm s} \rangle_{\mu} &=& \f{ \langle {\bm l}\cdot {\bm s} \rangle_{+} N_+(0) + \langle {\bm l}\cdot {\bm s} \rangle_{-} N_-(0)}{N_+(0) + N_-(0)} \nn \\
&=& 3\f{J_+^2 - J^2_-}{ 3J_+^2 + 2J^2_-}. \label{eq:sop}
\end{\eq}
$N_{\pm} (0)$ are the DOS for $J=2\pm 1/2$ state,
which are given by $N_{+} = \sum_{M=-5/2}^{5/2} {\rm Im} g^A_{+}(0)/\pi = \f{6}{\pi} 
{\rm Im} g^A_{+} (0) = 6N(0)\f{|V_d|^2}{(\mu - E^+)^2}$, and $N_{-} = 
\sum_{M=-3/2}^{3/2} {\rm Im} g^A_{-}(0)/\pi = \f{4}{\pi} {\rm Im} g^A_{-} (0) = 
4N(0)\f{|V_d|^2}{(\mu - E^-)^2}$, respectively.
Here,
$g^{A}_{\pm}(0)$ is the local Green function for $d$-electrons, which is given by
$g^{A}_{\pm} (0) = \f{1}{N}\sum_{\bk \sigma} \f{( V^{\pm}_{\bk M \sg})^{\ast}}{\mu - E^{\pm}} G^A_{\bk \sg}(0) \f{V^{\pm}_{\bk M' \sigma}}{\mu-E^{\pm}} = \f{|V_d|^2}{(\mu-E^{\pm})^2} \f{1}{N}\sum_{\bk} G^A_{\bk \sigma}(0) \delta_{M,M'}$.
From the above expression, note that $\ls$ vanishes when $J_+=J_-$.

\section{\label{3} $T$-matrix and Current Vertex Correction}

\begin{figure}[!htb]
\includegraphics[width=.85\linewidth, height=0.38\linewidth]{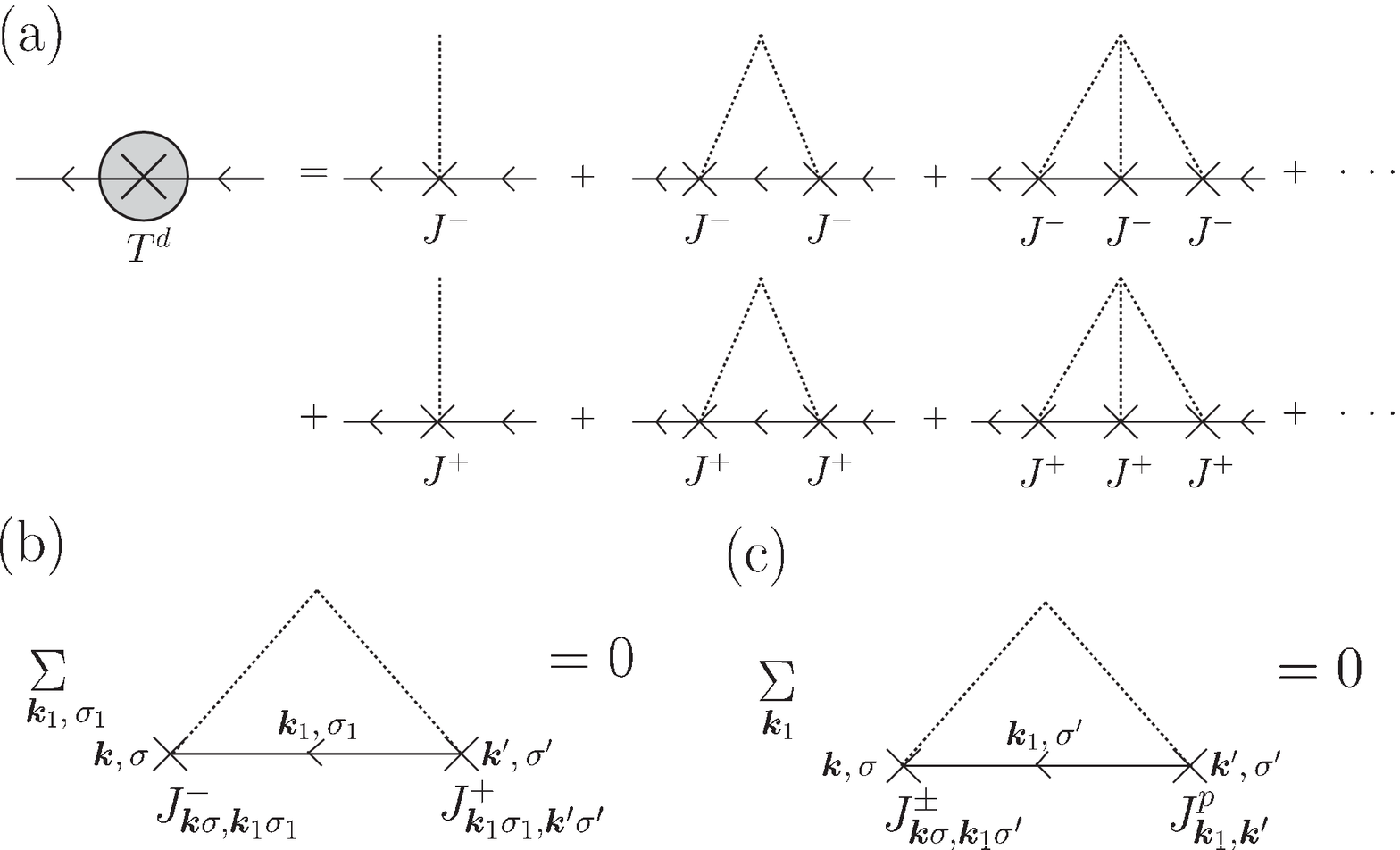} \\
\includegraphics[width=.7\linewidth, height=0.35\linewidth]{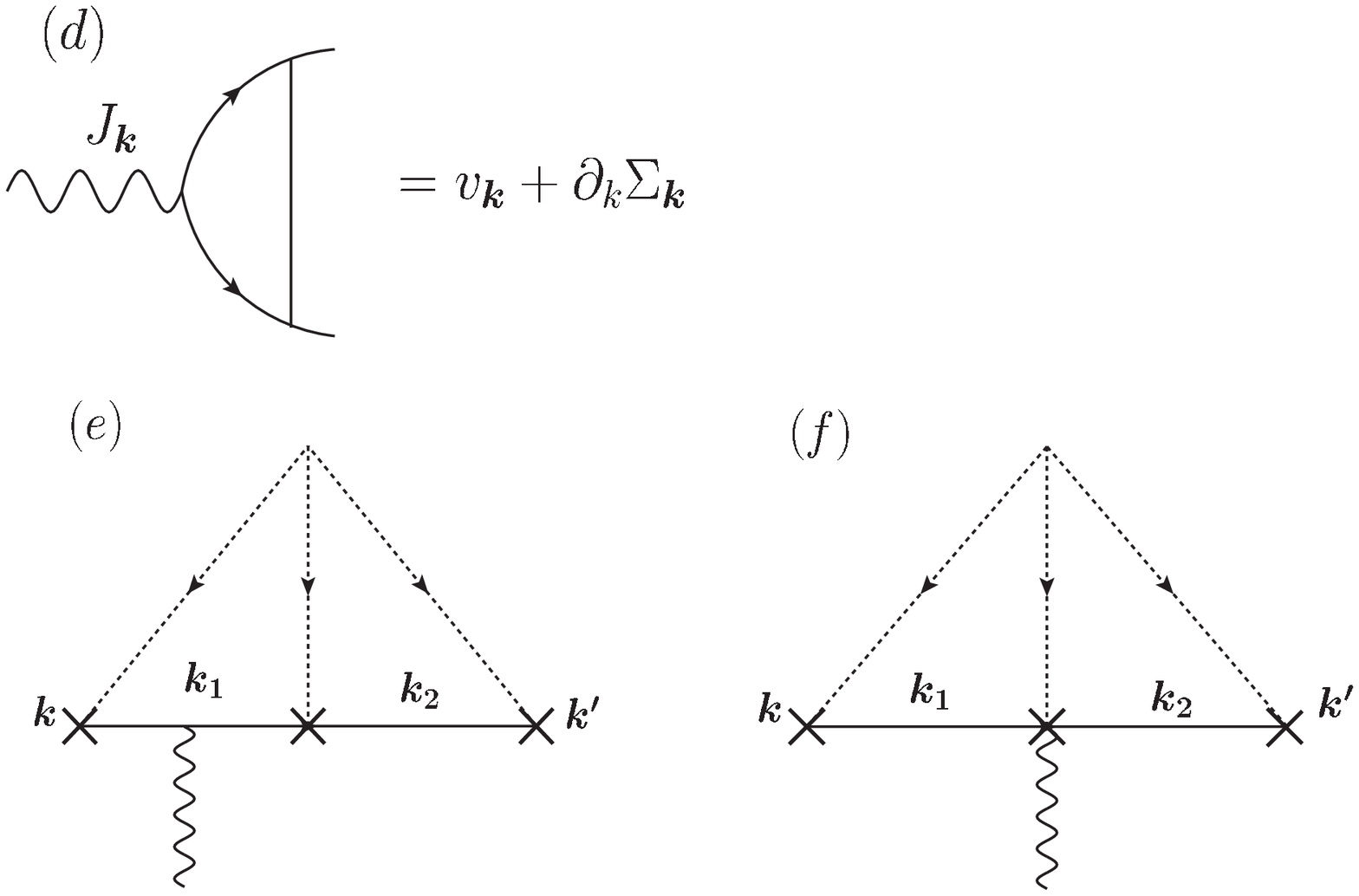}
\caption{\label{fig:tmat} (a) Diagrammatic expression for the $T$-matrix
due to $c$-$d$ resonant scattering. (b) and (c) Diagrams that vanish after $\bk_1$-summation.  
(d) Expression of the Ward-Takahashi identity connecting between the current vertex correction (CVC) and the self- energy $\it{\Sigma}_{\bm{k}}$.
(e) Type I of CVC is obtained by differentiating the Green function,
and (f) type II CVC obtained by differentiating the impurity potential.
The former (latter) process gives the skew scattering term (side jump term).
} 
\end{figure}

In this section, we study the scattering problem. 
In both models given by Eqs. (\ref{eq:Ham}) and (\ref{eq:Ham-re}),
the $T$-matrix due to the $c$-$d$ resonant scattering is equivalent;
it is given by
\begin{\eq}
T^d_{\bk\sg,\bk'\sg'} &=& J^{+}_{\bk\sg,\bk'\sg'} + J^{-}_{\bk\sg,\bk'\sg'}  \nn \\
&+& \frac{1}{N}\sum_{\bk_1,\sg_1} (J^{+}_{\bk\sg,\bk_1\sg_1} + J^{-}_{\bk\sg,\bk_1\sg_1}) G^0_{\bk_1} T^d_{\bk_1\sg_1,\bk'\sg'}, \label{eq:tmat}
\end{\eq}
where its diagrammatic expression is shown in Fig.\ref{fig:tmat} (a),
$N$ is the number of $\bk$-points,
and $G^0_{\bk} (\eps) =(\eps +\mu - \eps_{\bk})^{-1}$.
In this $T$-matrix, the terms containing both 
$J^{-}_{\bk\sg,\bk_1\sg_1}$ and $J^{+}_{\bk_1\sg_1,\bk'\sigma'}$ shown in Fig.\ref{fig:tmat} (b) 
vanish identically after $\bk_1$ and $\sigma_1$-summations
as follows:
\begin{\eq}
\sum_{\bk_1, \sg_1}J^{-}_{\bk\sg,\bk_1\sg_1} J^{+}_{\bk_1\sg_1,\bk'\sigma'} 
&\propto& \sum_{\sg_1} a^{M'-}_{m\sg_1} a^{M+}_{m\sg_1} \int d\Omega_{{\bm k_1}} Y^{M'}_{2} (\hat {\bk_1 }) \left( Y^{M}_{2} (\hat \bk_1) \right)^{\ast} \nn \\ 
&=& \delta_{M,M'} \sum_{\sg_1} a^{M-}_{m\sg_1} a^{M+}_{m\sg_1} =0. \label{eq:JJ-vanish}
\end{\eq}
%
%
Moreover, the term containing both $J^d_{\bk\sg,\bk'\sg'}$ and $J^p_{\bk,\bk'}$ given in Fig.\ref{fig:tmat} (c) vanishes identically due to 
the orthogonality of spherical harmonic functions.
%
%
%
Then, the solution of Eq. (\ref{eq:tmat}) for $\bk=\bk'$ and $\sg=\sg'$ is simply given by
\begin{\eq}
T^d_{\bk\sg,\bk\sg}(\eps) &=&  \f{ 2 J_{-} }{1-J_{-} g(\eps)} 
+  \f{ 3 J_{+}}{1-J_{+} g(\eps)},
\end{\eq}
where we have used the relations $\sum_{M} |V^{-(+)}_{\bk M \sg}|^2 =2 |V_{d}|^2 \ (3|V_d|^2)$, 
and $\frac{1}{N} \sum_{\bk \sg} V^{\pm}_{\bk M\sg} \\ \times G_{\bk}^0(\eps) \left( V^{\pm }_{\bk M'\sg} \right)^{\ast}=|V_d|^2 g(\eps) \delta_{MM'}$.
$g(\eps)=\frac{1}{N} \sum_{\bk} G^0_{\bk}(\eps)$ is the local Green function.
Assuming approximate particle-hole symmetry near $\mu$, we put 
$g^R(0)\equiv g(+i\delta)=-i \pi N(0)$.
%
Then, the quasiparticle damping rate in the $T$-matrix approximation 
is given by
\begin{\eq}
\g_d &=& -n_{\rm imp} {\rm Im} T^{dR}_{\bk\sg,\bk\sg}(0) \nn \\
&=& 2 n_{\rm imp} \f{\pi N(0) J_{-}^2}{1+ (\pi N(0) J_{-})^2} + 3 n_{\rm imp} \f{\pi N(0) J_{+}^2}{1+ (\pi N(0) J_{+})^2} \nn \\  
&\equiv& \gamma_{-} + \gamma_{+} \label{eq:gamma-T},
\end{\eq}
where $n_{\rm imp}$ is the impurity concentration.
Note that Eq. (\ref{eq:gamma-T}) is exact if $n_{\rm imp} \ll 1$.


Here, we discuss the current vertex correction (CVC) due to a single impurity,
which is the origin of the extrinsic terms.
The total current $J_{\bk}$ in the presence of the CVC is given by the Ward-Takahashi identity as follows,
which is shown in Fig.\ref{fig:tmat} (d) \cite{Nozieres}:
\begin{\eq}
J_{\bk} = v_{\bk} + \f{\rd \hat\Sigma_{\bk}}{\rd \bk},
\end{\eq}
where the second term in the right-hand side is the CVC,
which is required to satisfy the conservation law.
$\hat \Sigma_{\bk} = n_{\rm{imp}} \hat T_{\bk,\bk}$ is a self-energy in the T-matrix approximation.
As shown in Fig.\ref{fig:tmat}(e) and (f), two types of CVCs are derived from the Ward identity.
The type I of CVC is obtained by differentiating the Green function as shown in Fig.\ref{fig:tmat}(e),
and the type II of CVC is obtained by differentiating the impurity potential as shown in Fig.\ref{fig:tmat}(f).
We will show that the skew scattering term arises from the former type of the CVC
and the side jump term arises from the latter type.
In contrast, the CVC due to the local impurity potential is negligible in the intrinsic term in transition metals \cite{Tanaka-4d5d}.

\section{Skew Scattering Term \label{SS-term}}

\begin{figure}[!htb]
\includegraphics[width=.8\linewidth,height=.25\linewidth]{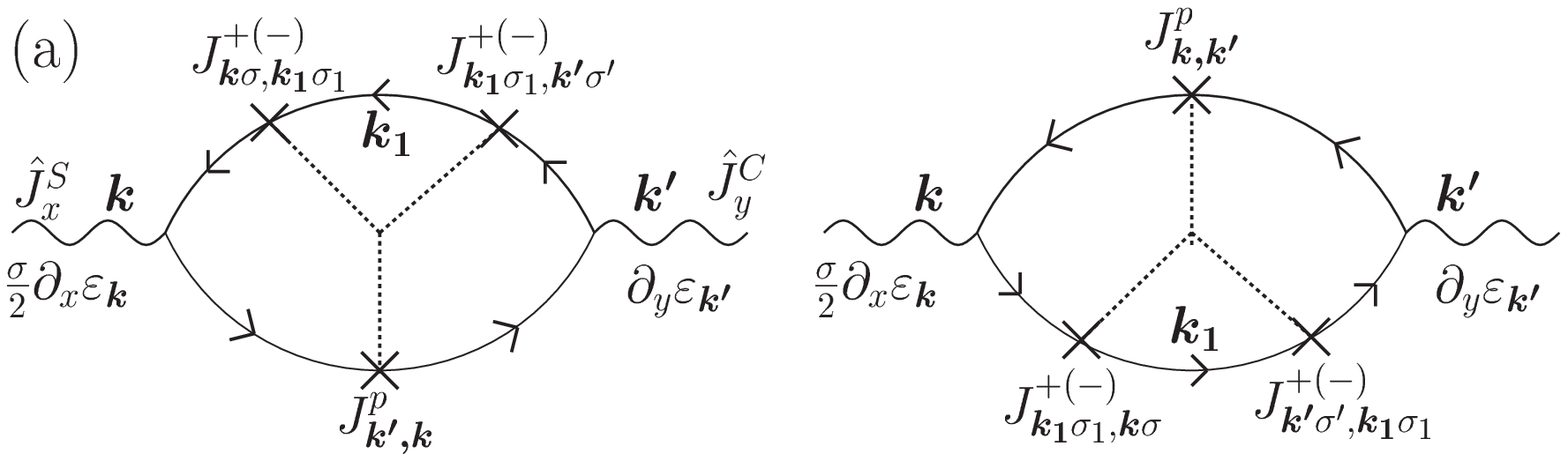} \\
\includegraphics[width=.8\linewidth,height=.25\linewidth]{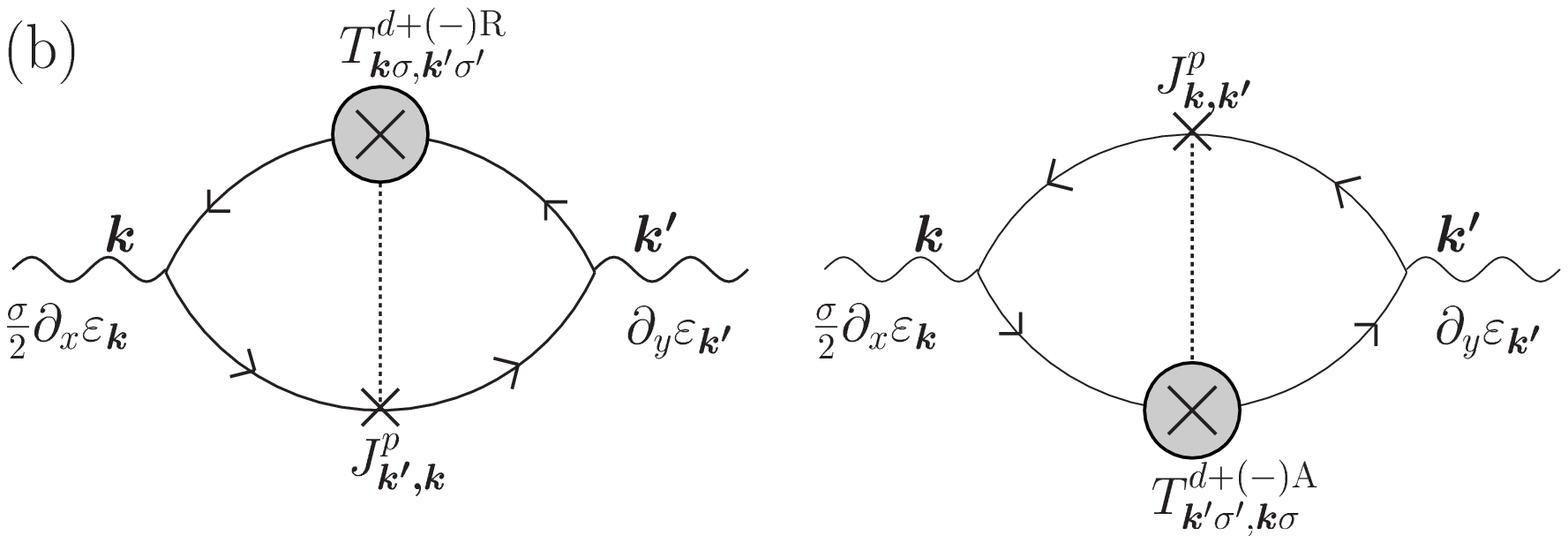}
\caption{\label{fig:ss-born} 
Diagrammatic expressions for SHC induced by
skew scattering (a) within the lowest order (extended Born 
approximation) contribution, and  
(b) in the $T$-matrix approximation with full order diagrams.
} 
\end{figure}

In this section, we study the skew scattering term using the linear response theory.
Initially, we consider the cases $N(0) |J_{\pm}| \ll 1$ 
and $|J_p| \ll |J_{\pm} |$, where the Born approximation is valid.
%
In analogy to Refs. \citen{Bruno} and \citen{Fert}, 
the lowest skew scattering term is given by the following second Born approximation,
\begin{\eq}
\sigma^{\rm ss (2nd Born)}_{\rm SH} &=& -\frac{e}{2\pi} n_{\rm imp} \frac{1}{N^2} \sum_{\bm{k,k'} 
\sigma } \frac{\sigma}{2} \frac{\partial 
\varepsilon_{\bm{k}}}{\partial k_x} \frac{\partial \varepsilon_{\bm{k'}}}{\partial k'_y} 
|G^R_{\bm{k}}(0)|^2  \nn \\
&\times& |G^R_{\bm{k'}}(0)|^2 \left\{ T^{d(2)R}_{\bk\sg,\bk'\sg}(0) J^p_{\bm{k',k}} + {\rm c.c.} \right\},  \label{eq:ss-born1}  
\end{\eq}
where $-e \ (e>0)$ is the electron charge. 
Its diagrammatic expression is shown in Fig.\ref{fig:ss-born} (a).
%
%
Here,
\begin{\eq}
T^{d (2)R}_{\bk\sg,\bk'\sg}(0) &=& \frac{1}{N} \sum_{\bk_1,\sg_1,\a=\pm} J^{\a}_{\bk\sg,\bk_1\sg_1} G^{R}_{\bm{k_1}}(0) J^{\a}_{\bk_1\sg_1,\bk'\sg} \nn \\
&=& \sum_{\a=\pm} g^R(0) J_{\a} J^{\a}_{\bk\sg,\bk'\sg}, 
\end{\eq}
is the second-order term of the $T$-matrix;
the first-order term in $J^{\pm}_{\bk\sg,\bk'\sg}$
does not contribute to $\sigma^{\rm ss}_{\rm SH}$
up to the first-order term in $J_{\pm}$ \cite{Fert}.
%
Note that any diagram that contains the 
part shown in Fig.\ref{fig:tmat} (b) vanishes identically.
The retarded Green function is given by
$G^{R}_{\bk}(0)=(\mu- \eps_{\bk} + i \g)^{-1}$, where $\gamma$ represents 
the quasiparticle damping rate.
Here, we put 
\begin{\eq}
\g=\g_d + \g_0, 
\end{\eq}
where $\g_0$ is the damping rate
due to nonmagnetic scattering, such as $c$-$p$ scattering ($\g_p = 3\pi 
n_{\rm imp}N(0) J^2_p$) and the scattering due to disorders. 
%
The charge current is given by $j^C_{\mu}=-e\rd \eps_{\bk}/\rd k_{\mu}$,
where $\mu=x,y$.
The spin current is then given by $j^S_{\mu} = (\sg/2) \rd \eps_{\bk}/\rd k_{\mu}$.

First, we consider the angular integration in Eq. (\ref{eq:ss-born1}),
which is given by
\begin{\eq}
\sum_{\sg} \f{\sg}{2} \left< J^{\pm}_{\bk\sg,\bk'\sg} J^p_{\bm{k',k}} f(\hat \bk, \hat \bk')  \right>_{\Omega},  \label{eq:average1}
\end{\eq}
where $\displaystyle f(\hat \bk, \hat \bk') \equiv \f{m^2}{k^2} \f{\rd \eps_{\bk}}{\rd k_x}\f{\rd \eps_{\bk'}}{\rd k'_y} = \sin \theta_{k} \cos \phi_{k} \sin \theta_{k'} \sin\phi_{k'} $,
and $\left< \cdot \cdot \cdot \right>_{\Omega}$ denotes the average over
the Fermi level, which is defined as $\displaystyle \left< A(\hat \bk, \hat \bk') \right>_{\Omega} \equiv \int \f{d\Omega_{k} d\Omega_{k'}}{(4\pi)^2} A(\hat \bk,  \hat \bk') $.
Since $f(\hat \bk, \hat \bk') =2\pi \left\{ Y^1_{1}(\hat \bk) - Y^{-1}_1(\hat \bk) \right\} \left\{ Y^{-1}_{1}(\hat \bk') + Y^{1}_1(\hat \bk') \right\}/3i $, angular integration such as
$\int d\Omega_{k} Y^{M-\sg/2}_{2}(\hat \bk)Y^{m}_{l}(\hat \bk)Y^{\pm1}_{1}(\hat 
\bk)$ appears in Eq. (\ref{eq:average1}). This integral is finite only when $l=1,3$.
Therefore, the interference of the $d \ (l=2)$ and $p \ (l=1)$ partial waves 
is essential for skew scattering \cite{Fert}.
After performing the angular integrations, Eq. (\ref{eq:average1})
is given by
\begin{\eq}
&&\sum_{\sg,\a=\pm} \f{\sg}{2} \left< J^{\a}_{\bk\sg,\bk'\sg} J^p_{\bm{k',k}} f(\hat \bk, \hat \bk')  \right>_{\Omega} 
 = - \f{i}{5} J_p ( J_{+} - J_{-}).
\end{\eq} 
%
Using the relations $|G^R_{\bk}(0)|^2 \approx \f{\pi}{\g} \delta(\mu-\eps_{\bk})$
for small $\g$, 
$\f{1}{N}\sum_{\bk_1} G^R_{\bk_1}(0)= g^R(0)= -i \pi N(0)$, and 
the spin-orbit polarization $\langle {\bm l}\cdot {\bm s}\rangle_{\mu}$ given in Eqs. (\ref{eq:sop}) and
(\ref{eq:ss-born1}) is transformed into
\begin{\eq}
\sxy^{\rm ss (2nd Born)} &=&  \frac{e}{2\pi } \frac{2 \pi^3 }{5} n_{\rm imp} J_p 
\f{\langle {\bm l}\cdot {\bm s} \rangle_{\mu}}{3} (3J^2_{+} + 2J^2_{-}) N(0)
\nn \\
&\times&\f{1}{N^2} \sum_{\bk,\bk'} \frac{\partial \varepsilon_k }{\partial k} 
\frac{\partial \varepsilon_k' }{\partial k'} \frac{1}{\g^2}
\delta(\mu-\varepsilon_k) \delta(\mu-\varepsilon_{k'}). \label{eq:ss-born2}
\end{\eq}
%
%
Since $\g_{-}= 2\pi n_{\rm imp} N(0) J_{-}^2$ ($\g_{+}= 3\pi n_{\rm imp} N(0) J_{+}^2$) for $J=3/2$ ($J=5/2$) in the Born approximation,
$\sxy^{\rm ss}$ is given by
\begin{\eq}
\sigma^{\rm ss(2nd Born)}_{\rm SH} = \frac{e}{2\pi} \frac{1}{30\pi^2} J_p k_{F}^4 \langle {\bm l}\cdot{\bm s} \rangle_{\mu} \f{\g_d}{\g^2}. \label{eq:ss-Born-32}
\end{\eq}
From the above expression, $\sxy^{\rm ss}$ 
vanishes when $J_+ = J_-$.


Now, we derive the skew scattering term 
using the $T$-matrix approximation, which gives the exact result for $n_{\rm imp} \ll 1$.
In this case, $T^{d(2) R}_{\bk\sg,\bk'\sg}(0)$ in Eq. (\ref{eq:ss-born1}) is replaced with the 
full $T$-matrix
$T^{dR}_{\bk\sg,\bk'\sg}(0)= \sum_{\a=\pm} T^{d(2) R}_{\bk\sg,\bk'\sg}(0) 
\left(1- g^R(0) J_{\a} \right)^{-1}$,
where the first-order term in $J^{\pm}_{\bk\sg,\bk'\sg}$ has been dropped.
The diagrammatic expression for $\sxy^{\rm ss}$ is shown in Fig.\ref{fig:ss-born} (b).
The angular integration in the $T$-matrix approximation
can be performed as 
\begin{\eq}
&&{\rm Re} \sum_{\sg} \f{\sg}{2} \left< T^{dR}_{\bk\sg,\bk'\sg}(0) J^p_{\bk',\bk}  f(\hat \bk, \hat \bk') \right>_{\Omega} \nn \\
&=&-\f{J_p}{5}  \left[ \f{ \pi N(0) J^2_{+}}{1+ \left( \pi N(0) J_{+} \right)^2} 
- \f{\pi N(0) J^2_{-}}{1+ \left( \pi N(0) J_{-} \right)^2} \right].
\end{\eq}
Then, the nonperturbative expression for $\sxy^{\rm ss}$ with respect to $J_{\pm}$
obtained by
the $T$-matrix approximation is given by
\begin{\eq}
\sigma_{\rm SH}^{\rm ss} &=& \f{e}{2\pi} \f{1}{30\pi^2}  J_p \f{k_F^4}{\g^2} \langle {\bm l}\cdot{\bm s} \rangle_{\mu} \nn \\
&\times& n_{\rm imp} \left\{  \f{ 3\pi N(0) J^2_+  + 2\pi N(0) J^2_-}{ \left[ 1+ (\pi N(0) J_+)^2 \right] \left[ 1+ (\pi N(0) J_-)^2 \right] } \right\}, \label{eq:ss-Tmat}
\end{\eq}
where $\ls$ is given in Eq.(\ref{eq:sop}).
In the case of $N(0) J_{+(-)} \gg 1$ and $N(0) J_{-(+)} \ll 1$, 
the quasiparticle damping rate in Eq. (\ref{eq:gamma-T}) is given as $\g_{+(-)}$. 
In the case of $\gamma_d = \gamma$ and $k_{\rm{F}}=\pi/a$, furthermore, the skew scattering term is simply given as
\begin{\eq}
\sigma_{\rm SH}^{\rm ss} = \f{e}{2 \pi a} \f{1}{10} n J_p \f{\ls}{\gamma}.
\end{\eq}
Here, $n=k^3_{\rm{F}}/3\pi^2 $ is the density of the conduction electrons.
Thus, $\sxy^{\rm{ss}} \propto \g_d/\g^2$
and therefore, considerably large SHC can be realized due to 
the skew scattering term $\sxy^{\rm{ss}}$ in the low resistivity metals.

%
%



\section{\label{sj-term} Side Jump Term I: for $\ls = -3/2$ or 1}
\label{sec:sj1}

\begin{figure}[!htb]
\includegraphics[width=.8\linewidth, height=0.2\linewidth]{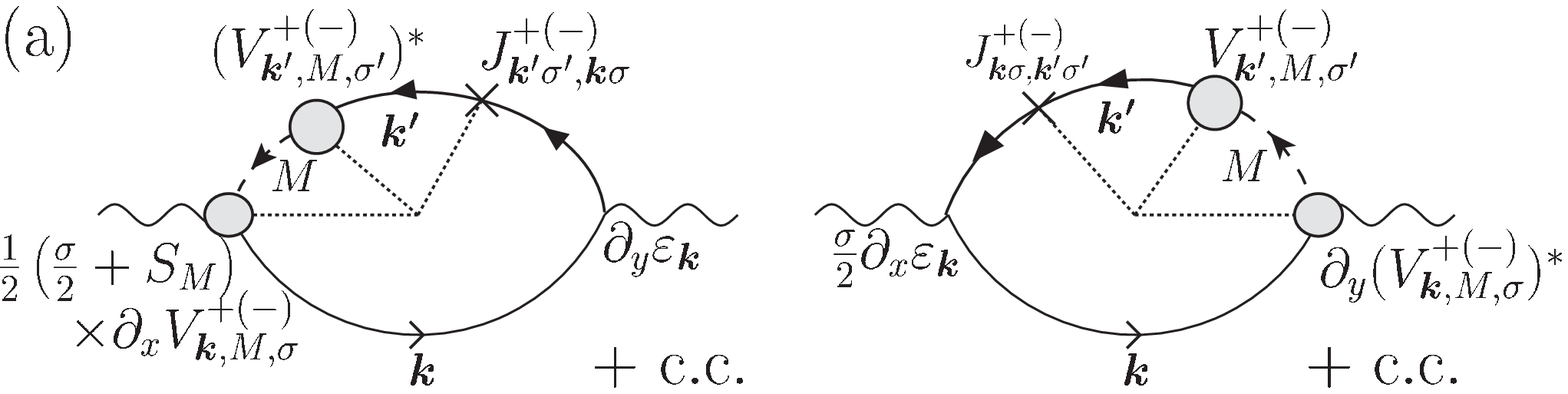} \\
\includegraphics[width=.8\linewidth, height=0.2\linewidth]{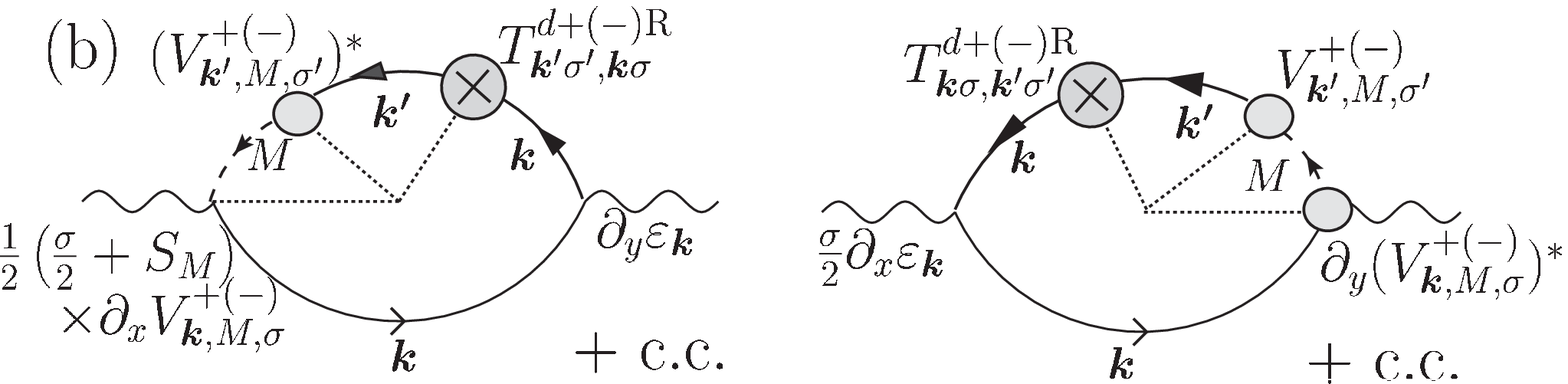} 
\caption{\label{fig:sj-born} (a) Diagrammatic expression for the side jump term
within the lowest order (Born approximation) contribution
and (b) in the $T$-matrix approximation. 
(c) The diagram without $J^{+(-)}$ vanishes identically. 
In contrast to the skew scattering term, $J_p$ is not necessary for the side jump.
(d) The diagram which vanishes identically unless $J_p$ is taken into account.
} 
\end{figure}

In this section, we derive the side jump term $\sxy^{\rm sj}$ for $J=5/2$ and $3/2$ independently
under the assumption that
the separation between these two states is complete ($J_{+} \gg J_{-}$ or $J_{-} \gg J_{+}$). 
However, when this separation is incomplete, the cross-term arises.
We will discuss the contribution of the cross term in the next subsection.
First, 
we consider the case where the Born approximation is valid:
$N(0)| J_{\pm}| \ll 1$.
In contrast to the skew scattering term, $J_p$ is not necessary for the side 
jump term.
%
The lowest order side jump term is given by the Born approximation.
In the present model, it is given by 
$\sxy^{\rm sj (Born)}$ is given by
\begin{\eq}
\sxy^{\rm sj (Born)} (J=5/2(3/2)) &=& 
-\f{e}{2\pi} n_{\rm{imp}}\f{1}{N^2} \nn 
\sum_{\substack{\bk,\bk' \\ \sg, \sg',  M}} \left\{ \f{1}{2}\left(\f{3\sg}{2}+S_M\right) \f{\rd V^{+(-)}_{\bk M \sg}}{\rd k_{x}}\right\} \f{\rd \eps_\bk}{\rd k_y} \nn \\
&\times& |G^R_{\bk}(0)|^2  
\f{1}{\mu-E^{+(-)}} \left[(V^{+(-)}_{\bk' M\sg'})^{\ast}G^R_{\bk'}(0)
J^{+(-)}_{\bk'\sg',\bk\sg} + \langle R  \leftrightarrow A \rangle \right], \nn \\ \label{eq:sj-born}
\end{\eq}
%
whose diagrammatic expression is shown in Fig.\ref{fig:sj-born}(a).
%
%
%
We stress that $J^{+(-)}$ is necessary for finite SHC
since the term without $J^{+(-)}$ vanishes identically as discussed in section \ref{sec:neg}.
In the present model, the charge current operator is given by $\hat j^C_{ \mu}=-e \hat v_{\bk \mu} $, where $-e \ (e>0)$ is the electron charge, and  
\begin{\eq}
\hat v_{\bk \mu} &=& \sum_{\sg} \f{\rd \eps_{\bk}}{\rd k_{\mu}}  c^{\dagger}_{\bk \sg}c_{\bk \sg} +
\sum_{\sg M} \left\{ \f{\rd V_{\bk M \sg}}{\rd k_{\mu}}  c^{\dagger}_{\bk \sg}d_{\bk M} + {\rm h.c.} \right\}. \label{eq:velocity}
\end{\eq}
Next, we explain the $s_z$-spin current operator $\hat j^S_{ \mu}$.
In the present model, $\hat s_z$ is given by
\begin{\eq}
\hat s_{z}&=& \sum_{\sg}\f{\sg}{2}c^{\dagger}_{\bk\sg}c_{\bk\sg} + \sum_{M} S^{-(+)}_{M}
(d^{-(+)}_{M})^{\dagger} d^{-(+)}_{M} 
+\sum_{M} S'_{M} (d^{-(+)}_M)^{\dagger} d^{+(-)}_M
\label{eq:sz},
\end{\eq}
where $S^{-(+)}_{M}$  is the spin $s_z$ for only the $J=3/2$ ($5/2$) state
given by
$S^{-(+)}_{M}=\sum_{\substack{m \sg}} \f{\sg}{2} \left[ a^{M-(+)}_{m\sg} \right]^2 $, 
and $S'_{M}$ is that of the cross-term for $J=3/2$ and 5/2 states given as 
$S'_{M}= \sum_{m \sg} \f{\sg}{2} a^{M-}_{m \sg}a^{M+}_{m \sg} $.
It is straightforward to show that 
$S^{-(+)}_{M}=-\f{M}{5}$ ($\f{M}{5}$) for $J=3/2$ $(5/2)$.
Moreover, $S'_{M}= -\f{1}{2}  \left\{ 1-\left( \f{2M}{5} \right)^2\right\}^{1/2} $ 
for both $J=3/2$ and $5/2$.
%
%
We now discuss the case where
$J=3/2$ and $5/2$ states are completely separated (i.e. $J_+ \gg J_-$ or $J_{-} \gg J _{+}$).
Then, $S'_M$ in Eq. (\ref{eq:sz}) is negligible, and 
the spin current 
$\hat j^S_{ \mu}\equiv\left\{ \hat v^c_{\bk \mu}, \hat s_z \right\}/2$ is given by
\begin{\eq}
\hat j^S_{ \mu} = \sum_{\sg}\f{\sg}{2} \f{\rd \eps_{\bk}}{\rd k_{\mu}} c^{\dagger}_{\bk\sg} c_{\bk \sg}
+  \sum_{\sg M}\left\{\f{1}{2} \left( \f{\sg}{2} + S_M \right) \f{\rd V^{\pm}_{\bk M \sg}}{\rd k_{\mu}} c^{\dagger}_{\bk \sg} d^{\pm}_{M} + \rm{h.c.} \right\}. \label{eq:spcurrent} 
\label{eq:spin-current}
\end{\eq}
Here, we study the velocity due to 
the $c$-$d$ mixing potential $V_{\bk M \sg}$ given as \cite{Kontani94}
\begin{\eq}
\f{\rd V^{\pm}_{\bk M \sg}}{\rd k_x} 
&=& -i \left( M-\f{\sg}{2} \right) \f{k_y}{k_x^2 + k_y^2} V^{\pm}_{\bk M \sg} 
+ \frac{\rd}{\rd k_x} \left(V^{\pm}_{\bk M \sg}\alpha_{M,\sg}^*\right)\alpha_{M,\sg}
 \nonumber \\
&\equiv& v_x^a + v_x^b
\label{eq:anomalous}.
\end{\eq}
$v_x^a$ is the anomalous velocity given by the $\bk$-derivative of 
the phase factor 
$\alpha_{M,\sg}=\rm{exp} \left\{i\left(M-\f{\sg}{2}\right)\phi_k\right\}$ 
in $V_{\bk M \sg}$ \cite{Kontani94,Tanaka-PAM}.
Since $v^a_x \propto k_y$ and thus 
$\sum_\bk v^a_x (\partial\epsilon_\bk/\partial k_y)\ne0$, 
the anomalous velocity gives rise to the large SHE and AHE 
in heavy fermion systems.
In contrast, $v^b_x \propto k_x$ gives a normal velocity.
%
%
Using the following relationships,
\begin{\eq}
\f{k_y}{k_x^2 + k_y^2} = \f{1}{k} \f{\sin\th \sin\phi}{\sin^2\th}, \ \f{\rd \eps_{\bk}}{\rd k_y} = \f{\rd \eps_{\bk}}{\rd k} \sin\th \sin\phi, \label{eq:rel1} 
\end{\eq}
Eq. (\ref{eq:sj-born}) is transformed as
\begin{\eq}
&& \sxy^{\rm{sj} (\rm{Born})} (J=5/2 (3/2))  \nn\\ 
&=&-\f{e}{2\pi} n_{\rm{imp}} \f{2}{\mu-E^{+(-)} } \sum_{\bk,\bk' \sigma, \sigma'}
 \f{\rd \eps_{\bk}}{\rd k} {\rm{Im}} G^R_{\bk'}(0) |G^R_{\bk}(0)|^2 \nn \\
&\times &  \left( \sum_{M} \Lambda^{+(-)}_{M \sg} V^{+(-)}_{\bk M \sg} (V^{+(-)}_{\bk' M \sg' })^{\ast}  \right) 
  J^{+(-)}_{\bk'\sg',\bk\sg} \sin^2 \phi_k ,  \label{eq:sj1}
\end{\eq}
where $\Lambda^{+(-)}_{M \sg}=\f{1}{2} \left(M-\f{\sg}{2} \right) \left( \f{3\sg}{2}+S^{+(-)}_M\right)$ and
$k \equiv |\bk|$. 
First, we derive $\sxy^{\rm{sj}}$ for $J=3/2$.
The angular integration in Eq. (\ref{eq:sj1}) is given by 
\begin{\eq}
&&\f{1}{\mu-E^-} \left< \sum_{\substack{M,\\ \sg,\sg'}} \Lambda^-_{M \sg} 
V^{-}_{\bk M \sg} (V^{-}_{\bk' M \sg'})^{\ast} J^{-}_{\bk'\sg',\bk\sg} \sin^2\phi_{k} \right \rangle_{\Omega}.  \label{eq:sj-anint-32}
\end{\eq}
Note that this term includes the term calculated as
$\sum_{\sg'}\left< (V^{-}_{\bk' M \sg'})^{\ast} V^{-}_{\bk'M'\sg'} \right>_{\Omega_{\bk'}} = |V_{-}|^2 \delta_{MM'}$.
Using the following relations for $J=3/2$,
\begin{\eq}
&&\sum_{M \sg} M^2 |V^{-}_{\bk M \sg}|^2 = |V_d|^2(1+6 \sin^2 \theta) , \\
&&\sum_{M \sg} \sg^2 |V^{-}_{\bk M \sg}|^2 = 4 |V_d|^2,  \\
&&\sum_{M \sg} M \sg |V^{-}_{\bk M \sg}|^2 = 2 |V_d|^2 \left( 1 - 3\sin^2\theta \right), \label{eq:Ms32}
\end{\eq}
the angular integration in Eq. (\ref{eq:sj-anint-32}) is performed as
\begin{\eq}
\f{|V_d|^2}{(\mu-E^-)^2}\left< \sum_{M \sg} \Lambda^{-}_{M \sg} 
|V^-_{\bk M \sg}|^2 \sin^2\phi_{k} \right>_{\Omega_{\bk} } \label{eq:sj-mid}
= -\f{18}{5} J_-^2.
\end{\eq}
Then, using the relations $|G^R_{\bk}|^2 \approx \f{\pi}{\gamma}\delta (\mu-\eps_{\bk}) $ for small $\gamma$, 
Eq. (\ref{eq:sj1}) is transformed as
\begin{\eq}
\sxy^{\rm{sj} (\rm{Born})} (J=3/2) = 
-\f{e}{2\pi}     \f{18}{5}  n_{\rm{imp}} J_-^2  \f{1}{N^2} \sum_{\bk \bk'} \f{1}{k}\f{\rd \eps_{\bk}}{\rd k} \f{\pi^2}{\gamma} \delta(\mu-\eps_{\bk})  \delta(\mu-\eps_{\bk'}).
\end{\eq}
By performing $\bk$ and $\bk'$-summations,
$\sxy^{\rm{sj}}$ for $J=3/2$ is given by
\begin{\eq}
\sxy^{\rm{sj} (\rm{Born})} (J=3/2)&=&-\f{e}{2\pi} n_{\rm{imp}}\f{9}{5}J_-^2 N(0) \f{k_F}{\gamma} \label{eq:sj-32-pre} \\
&=& \f{e}{2\pi a} \f{3}{5}\f{\gamma_-}{\gamma} \lsm. \label{eq:sj-3/2}
\end{\eq}
Here, $\lsm=-3/2$ and 
we put $k_F=\pi/a$, where $a$ is a lattice spacing.

In a similar way to the calculation of $\sxy^{\rm{sj}}$ for $J=3/2$, 
we derive $\sxy^{\rm{sj}}$ for $J=5/2$. 
In this case, using the following relations,
\begin{\eq}
&&\sum_{M \sg} M^2 |V^{+}_{\bk M \sg}|^2 = \f{3}{2}|V_d|^2 (1+16\sin^2 \theta) , \\
&&\sum_{M \sg} \sg^2 |V^{+}_{\bk M \sg}|^2 = 6 |V_d|^2,  \\
&&\sum_{M \sg} M \sg |V^{+}_{\bk M \sg}|^2 = 3 |V_d|^2 \left( 1 + 2\sin^2\theta \right), \label{eq:Ms52}
\end{\eq}
the angular integration is performed as
\begin{\eq}
&&\f{1}{\mu-E^+} \left< \sum_{\substack{M,\\ \sg,\sg'}} \Lambda^+_{M \sg} 
V^{+}_{\bk M \sg} (V^{+}_{\bk' M \sg'})^{\ast} J^{+}_{\bk'\sg',\bk\sg} \sin^2\phi_{k} \right>_{\Omega }  \label{eq:sj-anint-52} \nn \\
&=&\f{|V_{d}|^2}{(\mu-E^+)^2}\left< \sum_{M \sg} \Lambda^+_{M \sg} |V^+_{\bk M \sg}|^2 \sin^2 \phi_{\bk} \right>_{\Omega} 
=\f{11}{5} J_+^2. \label{eq:sj-anint-52}
\end{\eq}
%
Then,  by using $\lsp=1$ and $\gamma_+$, $\sxy^{\rm{sj}}$ for $J=5/2$ is given by
\begin{\eq}
\sxy^{\rm{sj} (\rm{Born} ) } (J=5/2) &=&\f{e}{2\pi} n_{\rm{imp}}\f{11}{5 } J^2_{+} N(0)\f{k_{\rm{F}}}{\gamma} \label{eq:sj-52-pre} \\
&=&\f{e}{2\pi a} \f{11}{15}\f{\gamma_+}{\gamma} \lsp, \label{eq:sj-5/2}
\end{\eq}
where we put $k_{{\rm{F}}}=\pi/a$.

Now, we derive the nonperturbative expression of the side jump term using the $T$-matrix approximation.
In this case, $J^{+(-)}_{\bk'\sg',\bk\sg}$ in Eq. (\ref{eq:sj-born}) is replaced by the following full T-matrix: 
\begin{\eq}
T^{dR+(-)}_{\bk\sg,\bk'\sg'}=\f{J^{+(-)}_{\bk\sg,\bk'\sg'}}{1-(g(0)J_{+(-)})^2}.
\end{\eq}
The diagrammatic expression for $\sxy^{\rm sj}$ is 
given in Fig.\ref{fig:sj-born} (b).
In this case, the angular integration in eqs. (\ref{eq:sj-anint-32}) and (\ref{eq:sj-anint-52}) is replaced as
\begin{\eq}
&&\f{1}{(\mu-E^{\pm})^2}\sum_{M \sg} \left< \Lambda_{M \sg} V^{\pm}_{\bk M \sg} (V^{\pm}_{\bk' M \sg'})^{\ast}
T^{dR\pm}_{\bk\sg,\bk'\sg'}  \sin^2\phi_{\bk} \right>_{\Omega} \nn \\
&=& -\f{18}{5}  \f{J^2_{-}}{1+(\pi N(0) J_{-})^2} \qquad \text{for} \ J=3/2, \\
&=&  \f{11}{5} \f{J^2_{+}}{1+(\pi N(0) J_{+})^2} \qquad \text{for} \ J=5/2.
\end{\eq}
Thus, $\sxy^{\rm{sj}}$ in the $T$-matrix approximation is given by $(1+(\pi N(0) J_{\pm})^2)$ times eqs. (\ref{eq:sj-32-pre}) and (\ref{eq:sj-52-pre}).
%
%
%
Therefore, the expression for the side jump term in 
eqs. (\ref{eq:sj-3/2}) and (\ref{eq:sj-5/2}) 
is valid beyond the Born approximation, by considering $\gamma_{\pm}$.




\section{\label{sj-term-cross} Side Jump term II: for general $\ls$}

In the previous section, we have derived 
$\sxy^{\rm{sj}}$ in the case 
where $J=3/2$ and $5/2$ states are completely isolated.
In this section, we discuss the case where these two states are not isolated,
where the cross-term between these two states appears.
Therefore, the contributions of these terms are indispensable for a more accurate calculation.
Note that the cross-term 
does not exist for the skew scattering term $\sxy^{\rm{ss}}$ because of the relationship in Eq. (\ref{eq:JJ-vanish}).
Now, we derive the cross-term for the side jump term $\sxy^{\rm{sj\mathchar`-cross}}$.
In the Born approximation, $\sxy^{\rm{sj\mathchar`-cross} \rm{(Born)}} $ is given by

\begin{figure}[!htp]
\includegraphics[width=.8\linewidth, height=0.2\linewidth]{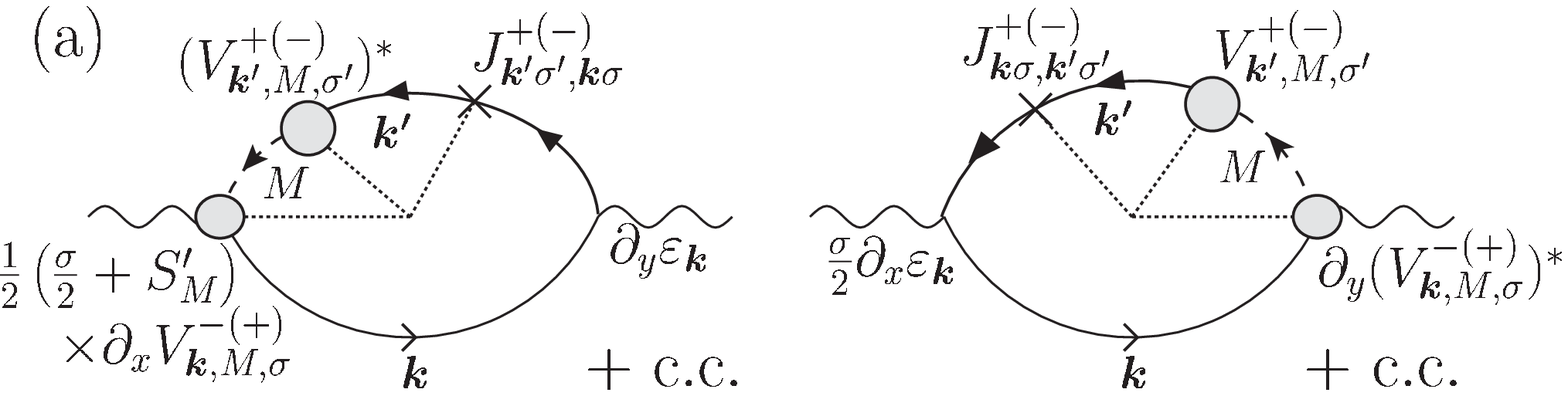} \\
\includegraphics[width=.8\linewidth, height=0.2\linewidth]{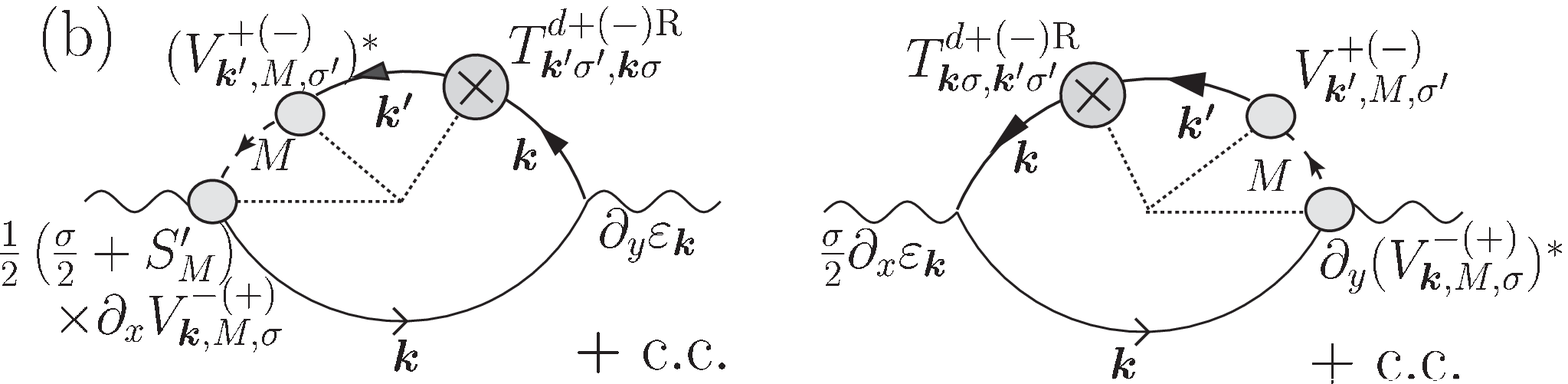}
\caption{\label{fig:sj-cross} 
Diagrammatic expression for the cross-term of 
the side jump term $\sxy^{\rm{sj\mathchar`-cross}}$(a)
within the lowest order (Born approximation) contribution,
and (b) in the $T$-matrix approximation. 
} 
\end{figure}


\begin{\eq}
\sxy^{\rm sj\mathchar`-cross (\rm{Born})} 
&=&-\f{e}{2\pi} n_{\rm{imp}}\f{1}{N^2} \sum_{\substack{\bk,\bk' \sg,\sg', \\ M,M',\alpha=\pm}} \left\{ \f{1}{2}\left(\f{3\sg}{2}+S'_M\right) \f{\rd V^{-\alpha}_{\bk M \sg}}{\rd k_{\mu}}\right\} \f{\rd \eps_\bk}{\rd k_y} \nn \\
&\times&  |G^R_{\bk}(0)|^2   \f{1}{\mu-E^{\alpha}} \left[(V^{\alpha}_{\bk' M\sg'})^{\ast}G^R_{\bk'}(0) J^{\alpha}_{\bk'\sg',\bk\sg} + \langle R  \leftrightarrow A \rangle \right], \label{eq:cross-+}
\end{\eq}
where $S'_{M}$ is given in Eq. (\ref{eq:sz}).
Its diagrammatic expression is shown in Fig.\ref{fig:sj-cross}(a). 
Here, the spin current operator is given by 
\begin{\eq}
\hat j^S_{ \mu} &=& \sum_{\sg}\f{\sg}{2} \f{\rd \eps_{\bk}}{\rd k_{\mu}} c^{\dagger}_{\bk\sg} c_{\bk \sg} \nn \\
&+&  \sum_{\sg M}\left\{\f{1}{2} \left( \f{\sg}{2} + S_M' \right) \f{\rd V^{\pm}_{\bk M \sg}}{\rd k_{\mu}} c^{\dagger}_{\bk \sg} d^{\pm}_{M} + \rm{h.c.} \right\}. \label{eq:spcurrent2} 
\end{\eq}
We remind the readers that the side jump term originates from 
the anomalous velocity due to the $c$-$d$ mixing potential. 
Using Eq. (\ref{eq:rel1}), 
Eq. (\ref{eq:cross-+}) is transformed as
\begin{\eq} 
\sxy^{\rm{sj\mathchar`-cross} \rm{(Born)}} &=& -\f{e}{2\pi} n_{\rm{imp}} 
 \sum_{\substack{\bk,\bk' \sigma, \sigma', \\ \alpha=\pm}}
\f{2}{\mu-E^{\alpha} }
 \f{\rd \eps_{\bk}}{\rd k} {\rm{Im}} G^R_{\bk'}(0) \nn \\
&\times& |G^R_{\bk}(0)|^2 
\left( \sum_{M} \Lambda'_{M \sg} V^{-\alpha}_{\bk M \sg} (V^{\alpha}_{\bk' M \sg' })^{\ast}  \right) \nn \\
 &\times& J^{\alpha}_{\bk'\sg',\bk\sg} \sin^2 \phi_k,   \label{eq:sj2} 
\end{\eq}
where $\Lambda^{'}_{M \sg}=\f{1}{2}(M-\f{\sg}{2})(\f{3\sg}{2} + S'_M) $.
As is the case with the previous subsection, 
the angular integration in Eq. (\ref{eq:cross-+})
is performed as
\begin{\eq}
&&\sum_{\alpha=\pm} \f{1}{\mu-E^{\alpha}} 
\left< \sum_{M \sg}\Lambda'_{M \sg} V^{-\alpha}_{\bk M \sg} (V^{\alpha}_{\bk' M \sg'})^{\ast}  J^{\alpha}_{\bk'\sg',\bk\sg} \sin^2 \phi_{\bk} \right>_{\Omega} \nn \\
&=& \sum_{\alpha=\pm} \f{1}{(\mu - E^{\alpha})^2} \left< \sum_{M\sg}\Lambda'_{M \sg} V^{-\alpha}_{\bk M \sg} (V^{\alpha}_{\bk M \sg})^{\ast} \sin^2 \phi_{\bk} \right>_{\Omega_{\bk}} \nn \\
&\times & \left< \sum_{M' \sg'} V^{\alpha}_{\bk'M' \sg'} (V^{\alpha}_{\bk' M \sg'})^{\ast} \right>_{\Omega_{\bk'}} \nn \\
&=& \sum_{\alpha=\pm} \f{|V_{d}|^2}{(\mu -E^{\alpha})^2}  \left< \sum_{M \sg} \Lambda'_{M \sg} V^{-\alpha}_{\bk M \sg} (V^{\alpha}_{\bk M \sg})^{\ast} \sin^2\phi_{\bk} \right>_{\Omega_{\bk}} \nn \\
&=& -\f{1}{5} ( J^2_{+} - J^2_{-} ). 
\end{\eq}
After taking the $\bk$-summation in Eq. ({\ref{eq:sj2}}), 
$\sxy^{\rm{sj\mathchar`-cross} \rm{(Born)}}$ is given by 
\begin{\eq}
\sxy^{\rm sj\mathchar`-cross \rm{(Born)}}=-\f{e}{2\pi}n_{\rm{imp}}\f{1}{5} \left( J^2_+ + J^2_- \right)N(0) \f{k_F}{\gamma}. \label{eq:sj-cr}
\end{\eq}
Therefore, according to eqs. (\ref{eq:sj-32-pre}), (\ref{eq:sj-52-pre}) and (\ref{eq:sj-cr}),
the final expression for the side jump term including the cross-term 
in the Born approximation is given as
\begin{\eq}
\sxy^{\rm sj (Born)}&=&\f{e}{2\pi}n_{imp}\left\{ \f{11}{5}J_+^2- \f{9}{5}J^2_- \right\} N(0)\f{k_F}{\gamma} \nn \\
&-& \f{e}{2\pi}n_{imp}\left\{ \f{1}{5}J_+^2+\f{1}{5}J^2_+ \right\} N(0)\f{k_F}{\gamma} \nn \\
&=& \f{e}{2\pi}\f{2}{3} \langle {\bm l} \cdot {\bm s} \rangle_{\mu}\f{\gamma_d}{\gamma}\f{k_F}{\pi}. \label{eq:sj-final}
\end{\eq}

Now, we derive $\sxy^{\rm{sj\mathchar`-cross}}$ in the $T$-matrix approximation.
In this case, $J^{+(-)}_{\bk'\sg',\bk\sg}$ in Eq. (\ref{eq:cross-+}) is 
replaced with the full $T$-matrix $T^{dR+(-)}_{\bk'\sg',\bk\sg}$.
The diagrammatic expression for $\sxy^{\rm{sj\mathchar`-cross}}$ is shown in Fig.\ref{fig:sj-cross}.
As is the case with the Born approximation,
the angular integration in the $T$-matrix approximation can be performed as
\begin{\eq}
&&\sum_{\alpha=\pm}\f{1}{\mu- E^{\alpha}} 
\left<  \sum_{M \sg} \Lambda'_{M \sg} V^{-\alpha}_{\bk M \sg} 
(V^{\alpha}_{\bk' M \sg'})^{\ast} T^{dR\alpha}_{\bk'\sg,\bk\sg} \sin^2 \phi_{\bk}  \right>_{\Omega} \nn \\
&=&\f{1}{5} \left(   \f{J^2_+}{1+(\pi N(0) J_+)^2}  + \f{J^2_-}{1+(\pi N(0) J_-)^2} \right).
\end{\eq}
%
%
%
Therefore, the nonperturbative expression for the side jump term in the full $T$-matrix approximation is given as
\begin{\eq}
\sxy^{\rm{sj}}&=&\f{e}{2\pi} \f{2}{3} \ls \f{k_F}{\pi} \f{1}{\gamma} 
 n_{\rm{imp}}\left\{  \f{ 3\pi N(0) J^2_+  + 2\pi N(0) J^2_-}{ \left[ 1+ (\pi N(0) J_+)^2 \right] \left[ 1+ (\pi N(0) J_-)^2 \right] } \right\}. \label{eq:sj-Tmat} 
\end{\eq}
In the case of $N(0) J_{+(-)} \gg 1$ and $N(0) J_{-(+)} \ll 1$, 
the quasiparticle damping rate in Eq. (\ref{eq:gamma-T}) is given as $\g_{+(-)}$. 
In the case of $\gamma_d = \gamma$ and $k_{\rm{F}} = \pi/a$,  the side jump term is given as
\begin{\eq}
\sigma_{\rm SH}^{\rm sj} = \f{e}{2 \pi a} \f{2}{3} \ls.
\end{\eq}
The relationship $\sigma_{\rm SH}^{\rm sj} \propto \ls$ is derived
only when the cross-term is taken into account correctly.



%
%
Here, we discuss the  
quasiparticle damping rate $\g$ dependence of 
$\sxy^{\rm{ss}}$ and $\sxy^{\rm{sj}}$.
From eqs. (\ref{eq:ss-Tmat}) and (\ref{eq:sj-Tmat}),
$\sxy^{\rm{ss}} \propto 1/\g_d$, $\sxy^{\rm{sj}} \propto (\g_d)^0$
for $\g_d \gg \g_0$.
When $\sigma_{xx}\propto1/\g_d$, 
the spin Hall angle due to the skew scattering mechanism
is independent of $\gamma_d$, and therefore, independent of
the impurity concentration.
This is consistent with the recent experiment by Niimi {\it{et al.}} \cite{Niimi}.
%

Finally, we discuss the ratio between $\sxy^{\rm{ss}}$ and $\sxy^{\rm{sj}}$. 
By using Eqs. (\ref{eq:ss-Tmat}) and (\ref{eq:sj-Tmat}), 
the ratio $\sxy^{\rm{sj}}/\sxy^{\rm{ss}}$ is given by
$\f{20}{3\pi}\f{\g}{J_p n}$, where $n = k_{\rm F}^3 /3\pi^2$ is the density of the conduction electrons.
Therefore, $\sxy^{\rm{sj}}$ in $\sxy$ exceeds $\sxy^{\rm{ss}}$ in samples with high resistivity.

\section{Negligible terms for the side jump term}
\label{sec:neg}

\begin{figure}[!htp]
\includegraphics[width=.4\linewidth, height=0.2\linewidth]{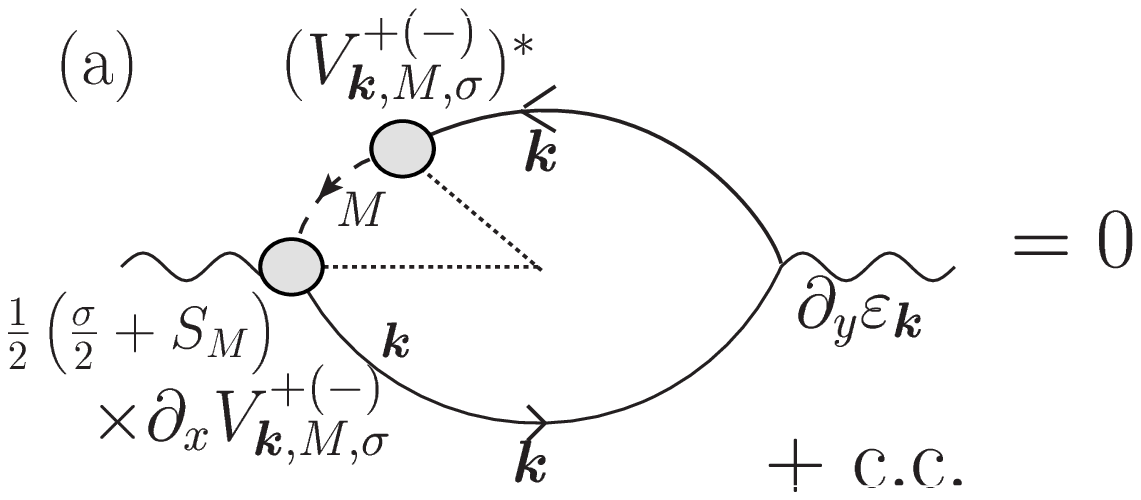} \ \
\includegraphics[width=.4\linewidth, height=0.2\linewidth]{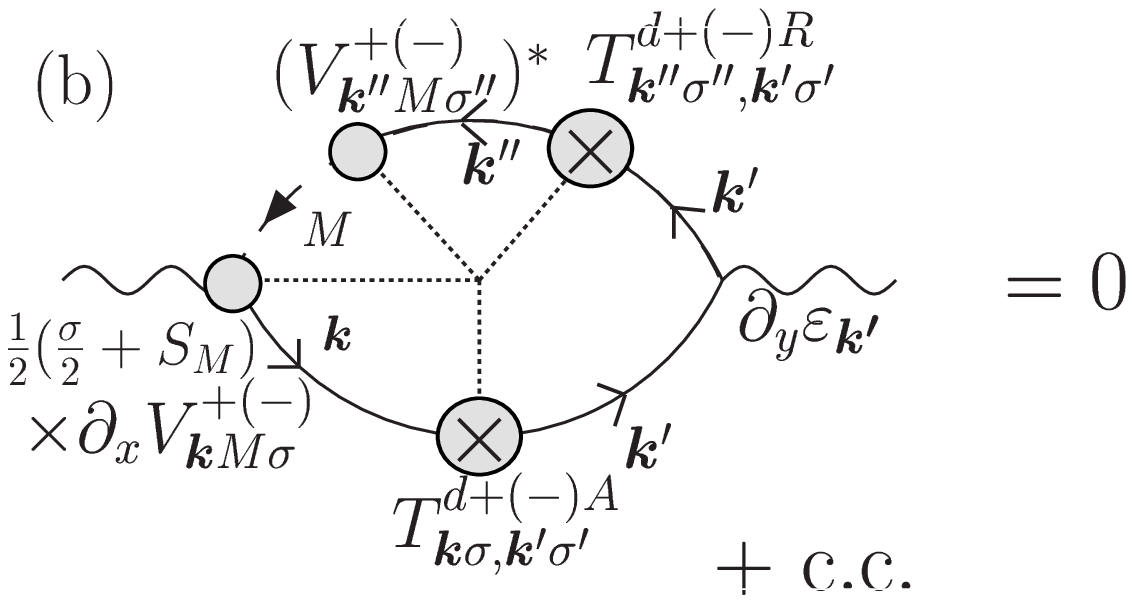} \\
\includegraphics[width=.8\linewidth, height=0.2\linewidth]{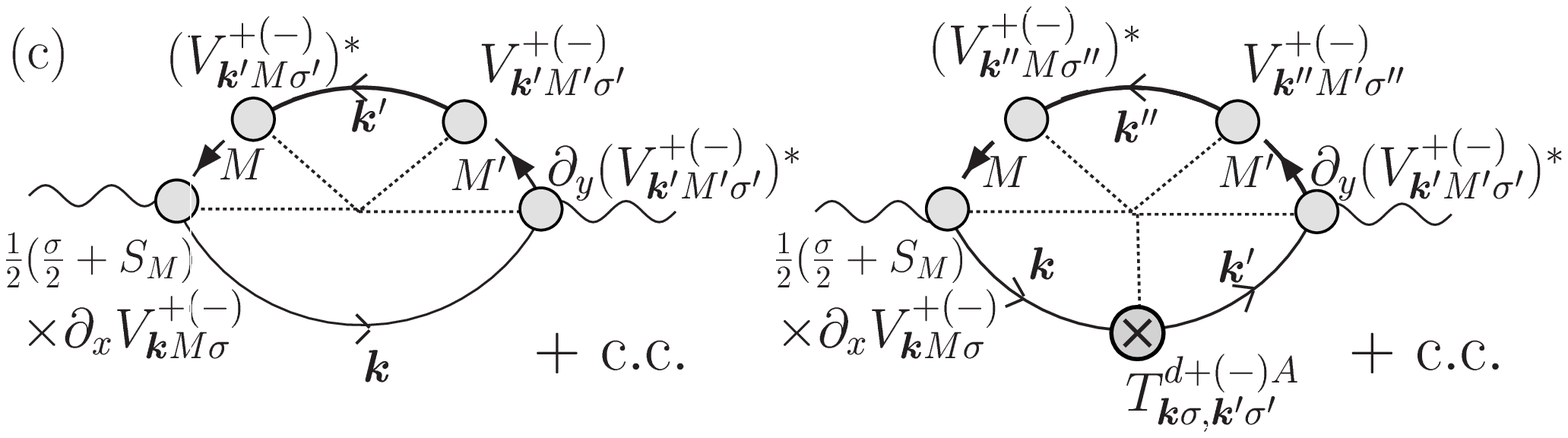}
\caption{\label{fig:sj-neg}
(a) The diagrammatic expression for the side jump term without $J^{+(-)}$ which vanishes identically. 
(b) The side jump term with both $T^{dR}_{\bk\sg,\bk'\sg'}$ and $T^{dA}_{\bk\sg,\bk'\sg'}$, 
which vanishes identically unless $J_p$ is taken into account.
(c) The side jump term with $\partial_{x} V$ and $\partial_{y} V$,
which are proportional to $n_{\rm{imp}}$.
Thus, they are negligible for $n_{\rm{imp}} \ll 1$.
} 
\end{figure}

Here, we discuss the terms for the side jump term, which vanishes identically or negligible.
In section \ref{sec:sj1},
we stress that $J^{+(-)}$ is necessary for a finite side jump term.
This is because the term without $J^{+(-)}$, which is shown in Fig. \ref{fig:sj-neg} (a),
vanishes identically since $\rm{Im} \frac{1}{\mu -E^{+(-)} + i \delta} =0$.
Furthermore, we also stress that the diagram shown in Fig. \ref{fig:sj-neg} (b) vanishes identically after 
$\bk'$-summation unless $J_p$ is taken into account.
At last,
the side jump term we have studied is given by the diagrams with $v^a_{\mu}$ in Eq. (\ref{eq:anomalous}) and $\f{\partial \eps_{\bk}}{\partial k_{\mu}}$.
We have dropped the diagrams with $v^a_{\mu}$ and $v^b_{\mu}$ as shown in Fig. \ref{fig:sj-neg} (c), since all of them are proportional to $n_{\rm{imp}}$, and therefore
negligible for $n_{\rm{imp}} \ll 1$.
In summary, the extrinsic SHE is given by the side jump term when $J_p =0$.
Then, the dominant term of the side jump term, which is of order $\gamma^0$ and $n^0_{\rm{imp}}$, 
is exactly given by Figs. \ref{fig:sj-born} and \ref{fig:sj-cross}.

\section{Discussion}

\subsection{The skew scattering mechanism in two-orbital model}

In the previous section \S \ref{SS-term}, we studied the SHC 
by using the Green function method. 
In this section, we discuss the skew scattering mechanism
based on the Boltzmann transport theory.
%
For this purpose, we study a simplified two-orbital model with $M=\pm 3/2$,
assuming a strong crystalline electric field.
In this model, 
\begin{\eq}
&&V^{-}_{\bk M \sg} \propto  -\sg V_d \left\{ \sqrt{4} Y^{-2\sg}_{2} (\hat \bk) \delta_{M,-3/2\sg} + Y^{\sg}_{2} (\hat \bk) \delta_{M,3/2\sg} \right\}, \nn \\
&&V^{+}_{\bk M \sg} \propto  V_d \left\{ \sqrt{4} Y^{2\sg}_{2} (\hat \bk) \delta_{M,3/2\sg} + Y^{-\sg}_{2} (\hat \bk) \delta_{M,-3/2\sg} \right\}. \nn
\end{\eq}
Since $V^{+ [-]}_{\bk M \sg} \propto Y^{2\sigma }_2 (\hat \bk) \ [Y^{-2\sigma }_2 (\hat \bk)] \propto {\rm e}^{2 i \sg \phi_{k}} \ [{\rm e}^{-2 i \sg \phi_{k}}]$ approximately,
the second-order term of $T^d$ is simply given as
\begin{\eq}
T^{d (2) R}_{\bk\sg,\bk'\sg} \sim -i\pi N(0) \left[ J_+^2 {\rm e}^{2i \sg 
\left( \phi_{\bk}-\phi_{\bk'} \right)} -\sg J^2_- {\rm e}^{-2i \sg 
\left( \phi_{\bk}-\phi_{\bk'} \right)}  \right].  \nn 
\end{\eq}
In the Boltzmann transport theory, the spin Hall resistivity due to skew 
scattering is
$\rho^{\rm ss}_{\rm SH} \propto \sum_{\sg} \sg \left< k_x k'_y w(\bk\sg \rightarrow \bk'\sg) \right>_{\mu}$, where 
$w$ represents the scattering probability,
which is proportional to $ \displaystyle n_{\rm imp} \left|T^{d(2) R}_{\bk\sg,\bk'\sg} + J^p_{\bk,\bk'} \right|^2$ due to Fermi's golden rule in the present model.
According to Ref. \citen{Fert},
skew scattering occurs when the scattering probability includes
an assymetric component 
$w^{\rm ss}(\bk \rightarrow \bk') \propto {\rm Im} {\rm e}^{i(\phi_k -\phi_{k'})} \propto (\hat \bk \times \hat \bk')_z$.
In the present model, $w^{\rm ss}$ arises from
the interference of the $d$ and $p$ scattering channel, 
$w^{\rm ss} \in \left( T^{d(2)R}_{\bk\sg,\bk'\sg} J^p_{\bk,\bk'} + {\rm c.c.} \right)$.
In fact, $ w^{\rm ss}(\bk\sg \rightarrow \bk'\sg) \propto (J^2_+ - J^2_-) \times {\rm Im} {\rm e}^{i\sg (\phi_k - \phi_{k'})} $ in this model 
since $J^p_{\bk,\bk'}$ contains the term
$Y^{\pm 1}_{1} (\hat \bk) \left[ Y^{\pm 1}_{1}(\hat \bk')  \right]^{\ast}  \propto {\rm e}^{\pm i (\phi_{k}-\phi_{k'})} $. 
In summary, a conduction electron with $\sg$ hybridizes with $l_z=-2\sg$
state due to the strong SOI, and therefore, the spin-dependent skew scattering 
probability $w^{\rm ss} (\bk\sigma \rightarrow \bk'\sigma) \propto (J^2_+ -J^2_-) {\rm Im} {\rm e}^{i\sg (\phi_k -\phi_{k'})}$ 
arises from the interference of the $d$ and $p$ angular momenta.
Note that the skew scattering contribution vanishes when $J_+ = J_-$.
Thus, the origin of the SHE due to the skew scattering mechanism is well understood
based on the simplified two-orbital model.

\subsection{Estimation of the spin Hall angle}

Here, we estimate the magnitude of the spin Hall angle due to
the skew scattering term: 
$\displaystyle \tan\aH \equiv \f{\sxy^{\rm ss}}{\sg_{xx}}\f{2e}{\hbar} $.
The longitudinal conductivity is given by
$\displaystyle \sg_{xx} = \f{e^2 n}{2 m \g}$,
where $n=k^3_F /3\pi^2$ is the density of the
conduction electrons.
Then, the spin Hall angle is given by
\begin{\eq}
\tan \alpha_{\rm SH} = \f{2\pi^2}{5} J_p N(0) \f{\g_d}{\g} \langle {\bm l} \cdot {\bm s} 
\rangle_{\mu}. \label{eq:Hallangle}
\end{\eq}
%
Since 
$\tan \delta_1 = -\pi N(0) J_p$ \cite{Hewson}, Eq. (\ref{eq:Hallangle}) can be 
rewritten as $\displaystyle \tan \alpha_{\rm SH} = -\f{2\pi}{5} \langle {\bm l} \cdot {\bm s} 
\rangle_{\mu} \f{\g_d}{\g} \delta_1$ if $|J_p| N(0) \ll 1$.

Next, we discuss the condition for a giant extrinsic SHE due to  
4$d$ and 5$d$ metal impurities.
From the expressions for the skew scattering term in Eq. (\ref{eq:ss-Tmat}) and 
the side jump term in Eq. (\ref{eq:sj-Tmat}),
the condition for the giant SHA is $\ls \sim O(1)$.
This seems difficult to be realized in $d$-electron systems
since $J=5/2$ and 3/2 states are energetically close, that is, $J_{+}\sim J_{-}$.
In Fe,
since the orbital magnetic moment is small,
the orbital degrees of freedom are quenched.
Therefore,
$\ls$ of Fe impurity in Au seems to be small \cite{Brewer,Costi}.
In contrast, the condition $\langle {\bm l} \cdot {\bm s} 
\rangle_{\mu} \sim O(1)$ is realized in Ce and Yb atoms
due to the almost complete separation between $J=3\pm1/2$ state,
and as a result, the giant SHE is expected to emerge
due to Ce and Yb impurities.

\subsection{The relationship between intrinsic and side jump terms}


In this section, we discuss the relationship between the intrinsic term $\sxy^{\rm{int}}$ and the side jump term $\sxy^{\rm{sj}}$.
Since these two terms originate from the anomalous velocity,
a close relationship between these two terms could be expected
\cite{Lyo}.
However, quantitative calculation that supports this idea
has not been performed.
To compare these two terms quantitatively, we compare 
the coefficient given as
$c^{\rm{int(sj)}}=\sxy^{\rm{int(sj)}} / \f{e}{2\pi a} \ls$, where $a$
represents the atomic distance.

According to Eq. (\ref{eq:sj-5/2}), the coefficient of the side jump term for $J=5/2$ is given by $c^{\rm{sj}}=\f{2}{3}$.
On the other hand, we have recently studied the intrinsic SHE based on the orbitally degenerate periodic Anderson model (OD-PAM) in Appendix \ref{App-A} \cite{Tanaka-PAM}.
The obtained intrinsic term for $J=5/2$ is given as
\begin{\eq}                                                                                                     
\sxy^{\rm{int}}&=&\f{e}{2\pi a}\f{11}{15}\ls \ \ \ \text{for} \ \ J=5/2, \label{eq:int-5/2}
\end{\eq}
where $\ls= 1$.
Therefore, the coefficient $c^{\rm{int}}$
for the OD-PAM is given by $c^{\rm{int}}=\f{11}{15}$.  
As a result, the coefficients $c^{\rm{int}}$ and $c^{\rm{sj}}$
obtained from OD-PAM and SIAM are comparable.
Furthermore, in the previous study,
we calculated $\sxy^{\rm{int}}$ and $\ls$
in various 4$d$ and 5$d$ transition metals based on a multiorbital tight-binding model
\cite{Tanaka-4d5d,Kontani-OHE}.
The coefficient $c^{\rm{int}}$ for Pt, for example, is estimated as $\sim 3$.
This is because $\sxy^{\rm{int}} \sim 10^{3}$ and $\ls \sim 0.33$ in Pt.
For the other transition metals, $c^{\rm{int}}$ is estimated to be of order 1.
%
Therefore, 
these two terms not only have the same origin,
but also have quantitatively comparable contributions.

%
%


Our study indicates that it might be difficult to distinguish the intrinsic 
mechanism from the side jump mechanism when the $d$-ions are strongly disordered.
For a more quantitative study, the study of both mechanisms
based on a realistic model is needed.
This is an important future problem.
Another future problem is to study the ac Hall conductivity.
Previously, the present authors studied the intrinsic ac anomalous Hall effect (AHE) \cite{Tanaka-AC}. 
Therein, the intrinsic ac AHC shows a prominent deviation from the Drude-type behavior.
This coincides with a recent experiment by Kim {\it{et al.}} \cite{Kim-AC}.
On the other hand, although no studies have reported the extrinsic ac AHE or SHE,
both the skew scattering term and side jump term may show 
a Drude-type behavior.
Therefore, ac SHE measurements will be useful to distinguish between the SHE
due to the intrinsic effect and that due to the extrinsic effect.
This is an important future work.




 


\subsection{Comparison with the study based on first principle calculation}
Recently, both the intrinsic and extrinsic SHEs have been studied based on 
the KKR Muffin-tin approximation \cite{Lowitzer}.
The calculated skew scattering term based on the Kubo formula 
agrees well with that based on the Boltzmann formula.
As for the side jump term, however, 
their obtained values seem to be much smaller than the side jump term based on the SIAM given in the present study.
This difference seems to be due to a different definition of the side jump term. 
In Ref. \citen{Lowitzer}, the type II CVC in Fig.\ref{fig:tmat}(f) seems to be dropped.
We expect that this is the reason why the side jump term obtained in their study is small in magnitude.

\subsection{Summary}
In summary, we have studied the extrinsic SHE due to transition metal impurities based on $J=2\pm1/2$ single-impurity Anderson model (SIAM). 
The analytical expressions for both the skew scattering term $\sxy^{\rm ss}$ and 
side jump term $\sxy^{\rm sj}$ have been derived. 
As is the case with the intrinsic term,
it was found that both the skew scattering and side jump terms are in proportion to $\ls$.
Therefore, 
$\ls \sim O(1)$ is a necessary condition for giant intrinsic and extrinsic SHE.
Moreover, we found a nontrivial close relationship between the intrinsic and side jump terms due to the fact that these two terms originate from the anomalous velocity.
We have shown that $C^{\rm{int}}=\sxy^{\rm{int}}/\ls$ in the OD-PAM 
is very close to
$C^{\rm{sj}}=\sxy^{\rm{sj}}/\ls$ due to dilute impurities described in SIAM.
Furthermore, according to the previous study \cite{Tanaka-4d5d}, 
$c^{\rm{int}}$ in the $4d5d$ transition metals is estimated to be of the same order.
f
To distinguish the intrinsic mechanism from the side jump mechanism,
the ac Hall effect would be useful \cite{Kim-AC}.



\section*{Acknowledgements}
We are grateful to 
D. S. Hirashima, J. Inoue, Y. Otani, Y. Niimi, and K. Yamada
for fruitful discussions.
One of the authors (T.T.) 
acknowledges the financial support
from the Japan Society for the Promotion of Science (JSPS).
This work was supported by a Grant-in-Aid for Scientific Research on
Innovative Areas "Heavy Electrons" (No. 20102008) of the MEXT, Japan.



\appendix

\section{\label{App-A} Derivation of the intrisic term for $J=3/2$ and $J=5/2$}
 
In the main text, we have discussed the close relationship between 
the intrinsic term $\sxy^{\rm{int}}$ and side jump term $\sxy^{\rm{sj} }$.
Here, we derive the intrinsic term for $J=3/2$ and $5/2$
based on the orbitally degenerate periodic Anderson model (OD-PAM) \cite{Tanaka-PAM}.

In the presence of the strong atomic SOI, the $J=5/2$ level is about 2000 K higher than the $J=3/2$ level. 
Therefore, we consider $J=5/2$ and \ 3/2 states separately.
We note that $\bm{l} \cdot \bm{s} =  \f{1}{2} \left[ j(j+1) -l(l+1) -s(s+1) \right] $ is given as
\begin{\eq}
\bm{l} \cdot \bm{s} = 1 \ \ \ {\rm{for}} \ \ J=5/2, \nn \\
\bm{l} \cdot \bm{s} = -3/2  \ \ \ {\rm{for}} \ \ J=3/2.
\end{\eq}
 
Here, we introduce the following OD-PAM Hamiltonian:
\begin{\eq}
\hat H&=&\sum_{\bk \sg} \eps_{\bk} c^{\dagger}_{\bk \sigma} c_{\bk\sg} +\sum_{\bk M} E^d d^{\dagger}_ {\bk M}d_{\bk M} \nn \\
&+& \sum_{M \bk \sigma}( V^{\ast}_{\bk M \sg} d^{\dagger}_{\bk M} c_{\bk \sg} 
+V_{\bk M \sg} c^{\dagger}_{\bk \sg} d_{\bk M}) \nn \\
&+&U\sum_{i,M\neq M'} n^d_{i M}n^d_{i M'},
\end{\eq}
where $c_{\bk \sg}^{\dagger}$ is the creation operator of a conduction electron with spin $\sg=\pm 1$.
$d^{\dagger}_{\bk M}$ is an operator of a $d$ electron with total angular momentum $J=5/2$ or 3/2 and 
$z$-component $M \ (-J\leq M\leq J)$.
$\eps_{\bk}$ is the energy for $c$ electrons,
$E^d$ is the localized $d$-level energy,
 and $U$ is the Coulomb interaction for $d$ electrons.
$V_{\bk M \sg}$ is the mixing potential between $c$ and $d$ electrons, which is given by
\begin{\eq}
V_{\bk M \sg}= \sqrt{\f{2}{2J+1}}\sqrt{4\pi} V_d \sum_{m} a^{M}_{m \sg} Y^{m}_{l}  (\theta_\bk,\phi_\bk),
\end{\eq}
where $a^{M}_{m \sg}$ is the Clebsch-Gordan (C-G) coefficient given by Eq. (\ref{eq:J5/2}) in the main text
and $Y^m_l (\th_k,\phi_k)$ is the spherical harmonic function.

Here, the conduction and $d$-electron Green's function for the OD-PAM in the absence of the magnetic 
field are given as \cite{Kontani94,Tanaka-PAM}
\begin{\eq}
G^{c}_{\bk \sg \sg}(\omega)&=& \left( \omega + \mu -\eps_{\bk} -\sum_{M} \f{|V_{\bk M \sg}|^2}{\omega+ \mu -E^d_M} \right)^{-1},  \nn \\
G^d_{\bk M  M'} (\omega) &=&G^{0 d}_{\bk M}(\omega) \delta_{M M'} \nn \\ 
 &+&\sum_{\sg} G^{0 d}_{\bk M}(\omega)V^{\ast}_{\bk M \sg} G^c_{\bk \sg}(\omega) V_{\bk M' \sg} G^{0 d}_{\bk M'}(\omega). 
\end{\eq} 
We note that $G^c_{\bk \sg \bar\sg}(\w)=0$ \ \cite{Kontani94} and 
$G^{0 d}_{\bk M}$ is the $d$-electron Green's function without hybridization given as 
\begin{\eq}
G^{0 d}_{\bk M}(\omega)=\f{1}{\omega + \mu -E^d}.
\end{\eq}

Now, we consider the quasiparticle damping rate $\hat \Gamma(\w)$, which is mainly 
given by the imaginary part of the $d$-electron self-energy $\it{\hat \Sigma} (\w)$.
In the present study, we assume that $\Gamma$ is diagonal with respect to $M$ $\Gamma_{MM'} = \gamma \delta_{MM'}$ and is independent of the momentum.
Here, we perform a calculation of the SHC using this constant 
$\g$ approximation. Then, the retarded (advanced) Green's function is given by
\begin{\eq}
G^{c \rm{R(A)}}_{\bk} (\w) &=&\left( \w +\mu -\eps_{\bk} - \f{|V_d|^2}{\w+\mu -E^d +(-) i \g}  \right)^{-1}, \nn \\
G^{0d \rm{R(A)}}_{\bk}(\omega) &=&\left[\w+\mu-E^d+(-) i \g \right]^{-1} .
\end{\eq}

We calculate the intrinsic SHC $\sxy^{\rm{int}}$ based on linear response theory.
The SHC at $T=0$ in the absence of the current vertex correction is given by  
$\sxy^{\rm{int}}= \sxy^{\rm{I}} + \sxy^{\rm{II}}$, where
\begin{\eq}
\sxy^{\rm{I}} &=& \f{1}{2\pi N} \sum_{\bk} \text{Tr} \left[ \hat j^{\rm{S}}_x \hat G^{\rm{R}} \hat j^{\rm{C}}_y \hat G^{\rm{A}} \right]_{\w=0}, \label{eq:FS-term} \\
\sxy^{\rm{II}} &=& \f{-1}{4\pi N} \sum_{\bk} \int_{-\infty}^{0} d\w \text{Tr} \left[ \hat j^{\rm{S}}_x \f{\rd \hat G^{\rm{R}}}{\rd \w} \hat j^{\rm{C}}_y \hat G^{\rm{R}} \right. \nn \\
&& \qquad \left. - \hat j^{\rm{S}}_x \hat G^{\rm{R}} \hat j^{\rm{C}}_y \f{\rd \hat G^{\rm{R}}}{\rd \w} -\langle \rm{R}\leftrightarrow \rm{A} \rangle \right]. \label{eq:sea-term}
\end{\eq}
Here,  $\sxy^{\rm{I}}$ and $\sxy^{\rm{II}}$ represent the Fermi surface  and Fermi sea term, respectively.

In the present model, the charge current operator is given by
$\hat j^C_{\mu}=-e\hat v_{\bk\mu}$, where $-e$ is the electron charge and
\begin{\eq}
\hat v_{\bk \mu}= \sum_{\sg} \f{\rd \eps_{\bk}}{\rd k_{\mu}}  c^{\dagger}_{\bk \sg} c_{\bk \sg} +
\sum_{\sg M} \left\{ \f{\rd V_{\bk M \sg}}{\rd k_{\mu}} c^{\dagger}_{\bk \sg} d_{\bk M} + {\rm{h.c.}} \right\}. \label{eq:velocity}
\end{\eq}
Next, we explain the $s_z$-spin current operator $\hat j^{s}_{\mu}$.
In the present model, $\hat s_z$ is given by 
\begin{\eq}
\hat s_z= \sum_{\sg} \f{\sg}{2} c^{\dagger}_{\bk \sg}c_{\bk\sg} + \sum_{M} S_{M} d^{\dagger}_{M \bk} d_{M \bk},
\end{\eq}
where $S_{M}=\sum_{m \sg} \f{\sg}{2} [a^M_{m \sg}]^2$.
It is straightforward to show that $S_M=-\f{M}{5} \ (\f{M}{5})$ for $J=3/2 \ (J=5/2)$.
Then, the spin current is given by
\begin{\eq}
\hat j^S_{\mu}&=& \sum_{\sg}\f{\sg}{2} \f{\rd \eps_{\bk}}{\rd k_{\mu}}c^{\dagger}_{\bk \sg}c_{\bk \sg} \nn \\
&+&\sum_{\sg M}\left\{ \f{1}{2}\left( \f{\sg}{2} + S_M \right) \f{\rd V_{\bk M \sg}}{\rd k_{\mu}} c^{\dagger}_{\bk \sg} d_{\bk M} + \rm{h.c.} \right\}. \label{eq:spincurrent}
\end{\eq} 
SHE originates from 
the anomalous velocity given by the $\bk$-derivative of the phase factor in $V_{\bk M \sg}$.
We note that the terms composed only of
$\rd_x \eps_{\bk} \rd_{y} \eps_{\bk}$ vanish identically \cite{Tanaka-PAM}.


 
Here, we calculate the SHC by neglecting CVC according to Eq. (\ref{eq:FS-term}), using eqs. (\ref{eq:velocity}) and (\ref{eq:spincurrent}).
$j^C_{\mu}$ and $j^S_{\mu}$ are composed of  the conduction electron term
$\rd \eps_{\bk}/\rd k_{\mu}  \equiv \rd_{\mu} \def \rd_{\mu} \eps_{\bk}$ and the hybridization term $\rd_{\mu} V_{\bk}$.
As studied in Ref. \citen{Tanaka-PAM}, the relationship
$\sxy^{\rm{int}} \approx \sxy^{\rm{I}}$ ( i.e. $\sxy^{\rm{I}} \gg \sxy^{\rm{II}}$)
is obtained for small $\gamma$.

According to Eqs. (\ref{eq:FS-term}), (\ref{eq:velocity}) and (\ref{eq:spincurrent}), the Fermi surface term
due to the product of $\partial_{\mu} V$ and $\partial_{\mu} \eps_{\bk}$ is given by
is given by
\begin{\eq}
\sxy^{\rm{I}} &=& \f{-e}{2\pi N} \sum_{\bk M \sg} \f{1}{2} \left( \f{3\sg}{2} + S_M \right) \nn \\
&\times & \left[  \f{\rd V_{\bk M \sg}}{\rd k_x} \f{\rd \eps_{\bk}}{\rd k_y} V^{\ast}_{\bk M \sg} |G^{cR}_{\bk}(0)|^2G^{0fR}_{\bk} (0) + {\rm c.c.} \right].
\label{eq:sxyI-1} 
\end{\eq}
%
%
%
First, we confine ourselves to the case for $J=5/2$ state.
Then, by using the following relationships
for $J=5/2$
\begin{\eq}
&&\sum_{M \sg} M^2 \V2 = \f{|V_d|^2}{2} \left( 1 + 16\sin^2\th \right), \label{eq:Ms1} \\
&&\sum_{M \sg} \sg^2 \V2 = 2 |V_d|^2,  \\
&&\sum_{M \sg} M \sg \V2 =  |V_d|^2 \left( 1 + 2\sin^2\th \right), \label{eq:Ms3} \\
&&\f{k_y}{k_x^2 + k_y^2} = \f{1}{k} \f{\sin\th \sin\phi}{\sin^2\th}, \ \f{\rd \eps_{\bk}}{\rd k_y} = \f{\rd \eps_{\bk}}{\rd k} \sin\th \sin\phi, 
\end{\eq}
Eq. (\ref{eq:sxyI-1}) is transformed as
\begin{\eq}
\sxy^{\rm{I}} = \f{e}{2\pi N} \f{22}{5} |V_d|^2 \sum_{k} \f{1}{k} \f{\rd \eps_{\bk}}{\rd k} \f{\g}{(\mu-E_{\bk})^2+\g^2} |G^{c R}_{\bk}(0)|^2, \label{eq:sxyI-2}  
\end{\eq}
where $k\equiv |\bk|$.

Here, we analyze Eq. (\ref{eq:sxyI-2}) when $\g$ is small enough. In this case,
\begin{\eq}
G^{cR}_{\bk}(0)&=& \left( \mu-\eps_{\bk} -\f{|V_d|^2}{\mu -E^d + i\g} \right)^{-1} \nn \\
 &\simeq & \left( \mu- \tilde \eps_{\bk} +i\G_c \right)^{-1},
\end{\eq}
where $\displaystyle \tilde \eps_{\bk} = \eps_{\bk} + \f{|V_d|^2}{\mu - E^d}$, and $\displaystyle \G_c= \f{|V_d|^2}{(\mu-E^d)^2}\g$.
Since $\gamma/(x^2+\gamma^2) = \pi \delta(x)$ for small $\gamma$,
we obtain the following relationship:
\begin{\eq}
|G^{cR}_{\bk}(0)|^2 &=& \f{1}{(\mu - E_{\bk})^2 + \G_c^2} \nn \\
&\simeq& \f{\pi}{\G_c} \delta(\mu-E_{\bk}).
\end{\eq}
%
%
Substituting the above equation into Eq. (\ref{eq:sxyI-2}), we obtain the 
following relationship for small $\g$:
\begin{\eq}
\sxy^{\rm{I}} &=& \f{e}{2\pi N } \f{22\pi}{15} \sum_{k} \f{1}{k} \f{\rd \eps_{\bk}}{\rd k} \delta(\mu-E_{\bk}). \label{eq:sxyI-3}
\end{\eq}
Now, we approximate the conduction electron as a free electron.
Then, $\sxy^{\rm{I}}$ for $J=5/2$ is given by
\begin{\eq}
\sxy^{\rm{I}}&=& e\f{11}{15} \f{k_F}{2\pi^2} N_{\rm{FS}} 
= \f{e}{2\pi a} \f{11}{15} N_{\rm{FS}}  \label{eq:I-va} \nn \\
&=&  \f{e}{2\pi a}\f{11}{15}N_{\rm{FS}}\ls,  \label{eq:int5/2}
\end{\eq}
where $\ls=1$, $a$ is the lattice spacing and
$N_{\rm{FS}}$ represents the number of large Fermi surfaces. 
The second line in Eq. (\ref{eq:int5/2}) is obtained by putting $k_{F}=\pi/a$.

At last, we derive the intrinsic term for $J=3/2$.
In this case, 
we use the following relationships:
\begin{\eq}
&&\sum_{M \sg} M^2 \V2 = \f{|V_d|^2}{2} \left( 1 + 6\sin^2\th \right), \label{eq:0Ms1} \\
&&\sum_{M \sg} \sg^2 \V2 = 2 |V_d|^2,  \\
&&\sum_{M \sg} M \sg \V2 =  |V_d|^2 \left( 1 - 3\sin^2\th \right). \label{eq:0Ms3} 
\end{\eq}
Then, we can perform the calculation of $\sxy^{\rm{int}}$
in a similar way to $J=5/2$.
As a result, $\sxy^{\rm{int}}$ for $J=3/2$ takes a negative value as
\begin{\eq}
\sxy^{\rm{I}}&=& -e\f{3}{5} \f{k_{F}}{2\pi^2} N_{\rm{FS}}= -\f{e}{2\pi a} \f{3}{5} N_{\rm{FS}} \nn \\
&=&\f{e}{2\pi a}\f{2}{5}N_{\rm{FS}}\ls. 
\end{\eq} 
Here, $\ls=-3/2$, and we put $k_{\rm{F}}=\pi/a$.


\section{\label{App-B} Accurate expression for the side jump term based on
a single-impurity Anderson model for rare-earth atoms}


In our previous study \cite{Tanaka-RE}, we derived the side jump term $\sxy^{\rm{sj}}$ based on the single-impurity Anderson model.
Therein, the definition of the spin current operator $\hat j^S_{\mu}$ was made by mistake.
Instead of defining 
\begin{\eq}
\hat j^S_{\mu}= \sum_{\sg}\f{\sg}{2} \f{\rd \eps_{\bk}}{\rd k_{\mu}} +\sum_{\sg M} \left\{ \f{\sg}{2} \f{\rd V^{f}_{\bk M \sg}}{\rd k_{\mu}} c^{\dagger}_{\bk \sg}f_M + \rm{h.c.} \right\}, \nn 
\end{\eq}
it should be defined by using the spin operator $\hat s_z$
as is the case with Eq. (\ref{eq:sz}) in the main text.
Then, the spin current operator is defined
as
\begin{\eq}
\hat j^S_{ \mu} &=& \sum_{\sg}\f{\sg}{2} \f{\rd \eps_{\bk}}{\rd k_{\mu}} c^{\dagger}_{\bk\sg} c_{\bk \sg} \nn \\
&+&  \sum_{\sg M}\left\{\f{1}{2} \left( \f{\sg}{2} + S_M \right) \f{\rd V^{f}_{\bk M \sg}}{\rd k_{\mu}} c^{\dagger}_{\bk \sg} f_{M} + \rm{h.c} \right\}.  
\end{\eq}
where $S_M= \f{\sg}{2} \left[ a^M_{m\sg}\right]^2$.
It is straightforward to show that $S_M=-\f{M}{7} \ (+\f{M}{7})$ for $J=5/2 \ (J=7/2)$.
By using the following relations for $J=5/2$,
\begin{\eq}
&&\sum_{M \sg} M^2 |V^{-}_{\bk M \sg}|^2 = \f{3}{2}|V_f|^2 (1+16\sin^2 \theta) , \\
&&\sum_{M \sg} \sg^2 |V^{-}_{\bk M \sg}|^2 = 6 |V_f|^2,  \\
&&\sum_{M \sg} M \sg |V^{-}_{\bk M \sg}|^2 = 3 |V_f|^2 \left( 1 -4\sin^2\theta \right), \label{eq:Ms52-RE}
\end{\eq}
and for $J=7/2$
\begin{\eq}
&&\sum_{M \sg} M^2 |V^{+}_{\bk M \sg}|^2 = 2|V_f|^2 (1+30\sin^2 \theta) , \\
&&\sum_{M \sg} \sg^2 |V^{+}_{\bk M \sg}|^2 = 8 |V_f|^2,  \\
&&\sum_{M \sg} M \sg |V^{+}_{\bk M \sg}|^2 = 4 |V_f|^2 \left( 1 + 3\sin^2\theta \right), \label{eq:Ms72-RE}
\end{\eq}
the final expression for $\sxy^{\rm{sj}}$ is given as
\begin{\eq}
\sxy^{\rm sj}&=& -\f{e}{2\pi}\f{6}{7}\f{k_F}{\pi} \f{\gamma_f}{\gamma} \nn \\
&=&\f{e}{2\pi} \f{3}{7} \langle {\bm l} \cdot {\bm s} \rangle_{\mu} \ \ \  {\rm{for}} \ J=5/2, \\
\sxy^{\rm sj} &=& \f{e}{2\pi}\f{15}{14}\f{k_F}{\pi} \f{\gamma_f}{\gamma} \nn \\
&=&\f{e}{2\pi} \f{5}{7} \langle {\bm l} \cdot {\bm s} \rangle_{\mu} \ \ \  {\rm{for}} \ J=7/2,
\end{\eq}
where spin-orbit polarization $\langle {\bm l} \cdot {\bm s} \rangle_{\mu}$ 
is given by $\langle {\bm l} \cdot {\bm s} \rangle_{\mu} =-2$ (3/2)
for $J=5/2$ ($J=7/2$).
We note that only the coefficient of the final expression
is slightly-modified compared with that in Ref. \citen{Tanaka-RE}.


\end{document}